\documentclass[twocolumn,tighten]{aastex61}

\newcommand{\ffdeg}{\mbox{\ensuremath{.\!\!\degr}}}
\newcommand{\ffarcs}{\mbox{\ensuremath{.\!\!^{\prime\prime}}}}

\begin{document}

\title{Variable dynamics in the inner disk of HD 135344B revealed with multi-epoch scattered light imaging\footnote{Based on observations collected at the European Organisation for Astronomical Research in the Southern Hemisphere under ESO programmes 087.C-0702(A,B), 087.C-0458(B,C), 087.C-0703(B), 088.C-0670(B), 088.D-0185(A), 088.C-0763(D), 089.C-0211(A), 091.C-0570(A), 095.C-0273(A), 097.C-0885(A), 097.C-0702(A), and 297.C-5023(A).}}

\correspondingauthor{Tomas Stolker}
\email{T.Stolker@uva.nl}

\author[0000-0002-5823-3072]{Tomas Stolker}
\affiliation{Anton Pannekoek Institute for Astronomy, University of Amsterdam, Science Park 904, 1098 XH Amsterdam, The Netherlands}

\author{Mike Sitko}
\affiliation{Department of Physics, University of Cincinnati, Cincinnati OH 45221, USA}

\author{Bernard Lazareff}
\affiliation{Universit\'{e} Grenoble Alpes, IPAG, F-38000 Grenoble, France CNRS, IPAG, F-38000 Grenoble, France}

\author{Myriam Benisty}
\affiliation{Universit\'{e} Grenoble Alpes, IPAG, F-38000 Grenoble, France CNRS, IPAG, F-38000 Grenoble, France}
\affiliation{Unidad Mixta Internacional Franco-Chilena de Astronoma (CNRS UMI 3386), Departamento de Astronom\'{i}a, Universidad de Chile, Camino El Observatorio 1515, Las Condes, Santiago, Chile}

\author{Carsten Dominik}
\affiliation{Anton Pannekoek Institute for Astronomy, University of Amsterdam, Science Park 904, 1098 XH Amsterdam, The Netherlands}

\author{Rens Waters}
\affiliation{Anton Pannekoek Institute for Astronomy, University of Amsterdam, Science Park 904, 1098 XH Amsterdam, The Netherlands}
\affiliation{SRON Netherlands Institute for Space Research, Sorbonnelaan 2, 3584 CA Utrecht, The Netherlands}

\author{Michiel Min}
\affiliation{Anton Pannekoek Institute for Astronomy, University of Amsterdam, Science Park 904, 1098 XH Amsterdam, The Netherlands}
\affiliation{SRON Netherlands Institute for Space Research, Sorbonnelaan 2, 3584 CA Utrecht, The Netherlands}

\author{Sebastian Perez}
\affiliation{Departamento de Astronom\'{i}a, Universidad de Chile, Casilla 36-D, Santiago, Chile}
\affiliation{Millennium Nucleus "Protoplanetary Disks," Chile}

\author{Julien Milli}
\affiliation{ESO, Alonso de C\'{o}rdova 3107, Vitacura, Casilla 19001, Santiago de Chile, Chile}

\author{Antonio Garufi}
\affiliation{Universidad Aut\'{o}noma de Madrid, Dpto. F\'{i}sica Te\'{o}rica, M\'{o}dulo 15, Facultad de Ciencias, Campus de Cantoblanco, E-28049 Madrid, Spain}

\author{Jozua de Boer}
\affiliation{Leiden Observatory, Leiden University, P.O. Box 9513, 2300 RA Leiden, The Netherlands}

\author{Christian Ginski}
\affiliation{Leiden Observatory, Leiden University, P.O. Box 9513, 2300 RA Leiden, The Netherlands}

\author{Stefan Kraus}
\affiliation{University of Exeter, Astrophysics Group, Stocker Road, Exeter, EX4 4QL, UK}

\author{Jean-Philippe Berger}
\affiliation{Universit\'{e} Grenoble Alpes, IPAG, F-38000 Grenoble, France CNRS, IPAG, F-38000 Grenoble, France}

\author{Henning Avenhaus}
\affiliation{Departamento de Astronom\'{i}a, Universidad de Chile, Casilla 36-D, Santiago, Chile}
\affiliation{Millennium Nucleus "Protoplanetary Disks", Chile}
\affiliation{Institute for Astronomy, ETH Zurich, Wolfgang-Pauli-Strasse 27, 8093 Zurich, Switzerland}

\begin{abstract}
We present multi-epoch Very Large Telescope/Spectro-Polarimetric High-contrast Exoplanet REsearch observations of the protoplanetary disk around HD~135344B (SAO~206462). The $J$-band scattered light imagery reveal, with high spatial resolution ($\sim$41\,mas, 6.4\,au), the disk surface beyond $\sim$20\,au. Temporal variations are identified in the azimuthal brightness distributions of all epochs, presumably related to the asymmetrically shading dust distribution in the inner disk. These shadows manifest themselves as narrow lanes, cast by localized density enhancements, and broader features which possibly trace the larger scale dynamics of the inner disk. We acquired visible and near-infrared photometry which shows variations up to 10\% in the $JHK$ bands, possibly correlated with the presence of the shadows. Analysis of archival Very Large Telescope Interferometer/PIONIER $H$-band visibilities constrain the orientation of the inner disk to $i = 18\ffdeg2^{+3.4}_{-4.1}$ and ${\rm PA} = 57\ffdeg3 \pm 5\ffdeg7$, consistent with an alignment with the outer disk or a minor disk warp of several degrees. The latter scenario could explain the broad, quasi-stationary shadowing in N-NW direction in case the inclination of the outer disk is slightly larger. The correlation between the shadowing and the near-infrared excess is quantified with a grid of radiative transfer models. The variability of the scattered light contrast requires extended variations in the inner disk atmosphere ($H/r \lesssim 0.2$). Possible mechanisms that may cause asymmetric variations in the optical depth ($\Delta\tau\lesssim1$) through the atmosphere of the inner disk include turbulent fluctuations, planetesimal collisions, or a dusty disk wind, possibly enhanced by a minor disk warp. A fine temporal sampling is required to follow day-to-day changes of the shadow patterns which may be a face-on variant of the UX~Orionis phenomenon.
\end{abstract}


\section{Introduction}

Spatially resolved observations provide detailed insight into the physical and chemical processes occurring in protoplanetary disks. The disk around the intermediate-mass star HD~135344B (SAO~206462) is a suitable target to be observed with high-resolution due to its proximity \citep[$156\pm11$\,pc;][]{gaia2016}, spatial extent \citep[$\sim$1\ffarcs15 in the scattered light;][]{grady2009}, low inclination \citep[$16\degr$;][]{vandermarel2016}, and brightness from visible to millimeter wavelengths \citep[e.g.,][]{carmona2014}. The disk is classified as a transition disk \citep[e.g.,][]{espaillat2014} with a large dust cavity resolved at continuum (sub)millimeter wavelengths \citep[$R_{\rm cav}=51$\,au at 156\,pc;][]{andrews2011}. In scattered light, two symmetric spiral arms have been detected \citep{muto2012}, which might indicate the presence of a massive gas giant in the outer disk \citep{dong2015a,fung2015,dong2017}. So far, only upper limits on planet masses have been derived from direct imaging observations \citep[3\,$M_{\rm Jup}$ at 0\ffarcs7, assuming hot-start evolutionary models;][]{maire2017}. The cavity radius in scattered light \citep[$R_{\rm cav}=27$\,au at 156\,pc;][]{stolker2016}, tracing micron-sized grains in the disk surface, is located inward with respect to the large grains in the midplane which can be explained by planet-induced dust filtration \citep{garufi2013}. The scattered light cavity coincides with a region in which the surface density of CO gas is significantly reduced \citep{vandermarel2016}.

Pre-main-sequence stars are commonly variable at optical and near-infrared wavelengths on various timescales, for example due to rotational modulation by stellar spots, variable accretion, dust obscuration, and structural changes in the inner disk \citep[e.g.,][]{eiroa2002}. Variability also occurs at mid-infrared wavelengths, for example, \emph{Spitzer}/IRS spectra show a typical anti-correlation between the amplitude of the near- and mid-infrared emission, indicating changes in the height of the inner disk at sub-au distance and consequent shadowing of the disk further outward \citep{espaillat2011}. The spectral energy distribution (SED) of HD~135344B contains a large near-infrared excess \citep[$F_{\rm NIR}/F_* = 0.27$;][]{garufi2017} due to the presence of hot dust in the innermost disk region \citep{brown2007}. It was noted by \citet{garufi2017} that HD~135344B belongs to a sub-category of group~I protoplanetary disks with both a large near-infrared excess and spiral arms, as well as shadows in most cases. The near-infrared continuum emission is variable up to 20--30\% while the 10\,$\mu$m flux exhibits fluctuations of 60\% \citep{grady2009,sitko2012}. Furthermore, \citet{grady2009} observed an anti-correlation between the strength of the $J$- and $L'$/$M'$-band fluxes which was linked to geometrical changes of the inner disk.

Multi-wavelength polarimetric differential imaging observations by \citet{stolker2016} revealed three shadow lanes in the $J$ band and a broader shaded region bound by two of the shadow lanes. A comparison with optical images from a month earlier showed that the southern $J$-band shadow lane was not present in the $RI$ bands, pointing toward a transient or variable origin. Those shadows are presumably cast by dust in the inner disk which is asymmetrically perturbed and/or misaligned with respect to the outer disk \citep{stolker2016}. Similarly, \citet{wisniewski2008} found that the scattered light flux from the protoplanetary disk around HD~163296 showed variations between different imagery epochs obtained with the \emph{Hubble Space Telescope (HST)}. This might indicate a time-variable shadowing of the outer disk by scale height variations of the inner disk wall \citep{wisniewski2008}, in line with the photometric variability in the near-infrared due to structural disk changes near the dust sublimation zone \citep{sitko2008}. Furthermore, \citet{ellerbroek2014} reported enhanced extinction in the optical for HD~163296, lasting from a few days up to a year, which was interpreted as caused by a dusty disk wind.

In this paper, we present multi-epoch, polarized scattered light imagery of the protoplanetary disk around HD~135344B that were obtained with the Spectro-Polarimetric High-contrast Exoplanet REsearch \citep[SPHERE;][]{beuzit2008} instrument. We aim to detect and characterize brightness variations caused by shading dust in the inner disk. The shadow patterns and their variability allow us probe to the physical processes occurring in the innermost disk region which are not directly accessible by high-contrast imaging instruments. The scattered light images are complemented with multi-epoch visible and near-infrared photometry, that we aim to link to the scattered light variations, and near-infrared interferometry, allowing us to place constraints on the orientation of the inner disk. Furthermore, we will use radiative transfer simulations to quantify the correlation between the scattered light contrast and near-infrared excess in order to estimate the extent of the inner disk variations.

\section{Observations and data reduction}\label{sec:observations}

\subsection{SPHERE/IRDIS dual-polarimetric imaging (DPI)}\label{sec:irdis}

\begin{deluxetable*}{lccccccc}
\tablecaption{SPHERE/IRDIS observations\label{tab:observations}}
\tabletypesize{\footnotesize}
\tablecolumns{8}
\tablewidth{0pt}
\tablehead{
\colhead{UT date} & \colhead{Integration\tablenotemark{a}} & \colhead{Airmass} & \colhead{Seeing\tablenotemark{b}} & \colhead{Wind speed\tablenotemark{c}} & \colhead{Coherence time\tablenotemark{d}} & \colhead{Strehl ratio\tablenotemark{e}} & \colhead{PSF FWHM\tablenotemark{f}} \\
\colhead{} & \colhead{(minutes)} & \colhead{} & \colhead{(arcsec)} & \colhead{(m\,s$^{-1}$)} & \colhead{(ms)} & \colhead{(\%)} & \colhead{(mas)}
}
\startdata
2015 May 03\tablenotemark{g} & \phn76.8 & 1.11--1.37 & 0.69(0.07) & \phn6.1(0.3) & \phn2.4(0.4) & 73(4) & $41.3 \times 38.4$  \\
2016 May 04                  &    102.4 & 1.06--1.38 & 0.52(0.16) & \phn1.4(0.7) &    13.1(3.4) & 78(2) & $42.0 \times 41.9$ \\
2016 May 12                  & \phn17.1 & 1.14--1.18 & 2.14(0.30) &    14.1(0.8) & \phn3.0(0.6) & 60(7) & $43.1 \times 44.8$  \\
2016 June 22                 & \phn34.1 & 1.09--1.15 & 0.63(0.07) & \phn8.1(0.3) & \phn8.3(1.4) & 72(2) & $40.1 \times 43.6$  \\
2016 June 30                 & \phn34.1 & 1.02--1.03 & 0.37(0.05) & \phn8.0(0.5) & \phn9.5(1.5) & 79(1) & $40.2 \times 38.1$  \\
\enddata
\tablecomments{Values in parenthesis provide the standard deviation of the average measured value.}
\tablenotetext{a}{Multiplication of the integration time (DIT), the number of integrations (NDIT), the number of polarimetric cycles (NPOL), and the number of half-wave plate orientations (NHWP=4).}
\tablenotetext{b}{Seeing measured by the differential image motion monitor (DIMM) at 0.5\,$\mu$m}
\tablenotetext{c}{Wind speed at ground level.}
\tablenotetext{d}{Coherence time measured by the multi-aperture scintillation sensor (MASS), except the first epoch which is estimated from the DIMM.}
\tablenotetext{e}{$H$-band Strehl ratio estimated by SAXO.}
\tablenotetext{f}{Point spread function full width at half maximum fitted to the non-coronagraphic flux frames.}
\tablenotetext{g}{Archival data from \citet{stolker2016}.}
\end{deluxetable*}

Imaging polarimetry data sets were obtained on 2016 May 03, 2016 May 11, 2016 June 22, and 2016 June 29 with the near-infrared imager \citep[IRDIS;][]{dohlen2008} of SPHERE at the European Southern Observatory's Very Large Telescope (VLT). Observations were carried out with the broadband $J$~filter (\texttt{BB\_J}, 1.245\,$\mu$m) in DPI \citep{langlois2014} mode. The pixel scale of the detector is 12.26\,mas\,pix$^{-1}$ \citep{maire2016}. An apodized Lyot coronagraph was employed (\texttt{N\_ALC\_YJH\_S}, 185\,mas mask diameter), allowing for an integration time of 32\,s. The four standard half-wave plate orientations were cycled with 2 or 4 subsequent integrations per half-wave plate orientation. The extreme adaptive optics system \citep[SAXO;][]{fusco2006} provided a typical Strehl ratio of $\sim$75\% in the $H$ band (see Table~\ref{tab:observations}).

Seeing conditions were mostly good ($<$0\ffarcs7) except on 2016 May 11 when the observations were executed with an average seeing of 2\ffarcs1 and the presence of strong winds. Nonetheless, the AO loop remained closed for a total integration time of 17\,minutes. The integration time was 34\,minutes or longer for the other observations. The point spread function (PSF) of the sequence on 2016 May 03 was partly affected by low-order aberrations caused by low wind speeds (1.4\,m\,s$^{-1}$ on average). The PSF quality was evaluated from the non-coronagraphic images recorded by the differential tip-tilt sensor after which 13 polarimetric cycles were removed, leaving a total of 11 cycles (47\,minutes).

Standard calibration procedures were applied with the SPHERE data reduction and handling (DRH) pipeline \citep[v0.18.0;][]{pavlov2008} which included sky subtraction, flat field correction, and bad pixel interpolation. The frames with the horizontally and vertically polarized flux were separated and subsequently processed by a custom pipeline for DPI data. We obtained coronagraphic images with four symmetric satellite spots, induced by a periodic modulation applied to the deformable mirror, before and after the science sequence. These frames were used to determine the position of the star behind the coronagraph mask. To center the coronagraphic DPI data, we interpolated linearly between the start and end position of the star. Stokes~$Q$ and $U$ images were obtained with the double-difference method \citep{hinkley2009} and subsequently collapsed with a mean stacking. 

To correct for instrumental polarization, we use the method described by \citet{canovas2011} which assumes that the central star is unpolarized. The $U_\phi$ signal, which provides an estimate of the noise level in the single-scattering limit, was minimized by stepwise changing the inner and outer radius of the annulus used to measure the (assumed to be unpolarized) signal close to the star. The azimuthal counterparts of the Stokes $Q$ and $U$ images, $Q_\phi$ and $U_\phi$, were calculated with an additional minimization applied on the $U_\phi$ image by correcting for a minor rotational offset of the half-wave plate \citep{avenhaus2014} for which the optimized values ranged from --2\ffdeg0 to --1\ffdeg3. We note that the procedure of minimizing the $U_\phi$ signal is not strictly valid because part of the scattered light flux from the disk will be present in the $U_\phi$ image due to multiple scattering \citep{canovas2015}. The effect will be small when the disk inclination is low, however, the high signal-to-noise ratio (S/N) of the disk detection around HD~135344B might reveal a real signal in the $U_\phi$ image \citep{stolker2016}. Finally, the images are rotated by --1\ffdeg8 toward the true north orientation \citep{maire2016}.

Flux frames were obtained at the start of each sequence, by shifting the star away from the coronagraph, with a shorter integration of 0.87\,s and an additional neutral density filter (\texttt{ND1.0}) in the optical path to avoid saturation. The flux frames are used to determine the angular resolution of the images (see Table~\ref{tab:observations}), as well as the scattered light contrast of the disk (see Section~\ref{sec:imagery}). The data reduction procedure included a dark frame subtraction, flat field correction, and bad pixel interpolation. To remove any residual background and bias from the images, we calculated for each detector column the mean pixel value (with the inner 1\ffarcs5 masked) and subtracted that value from all pixels in that column. The PSF of HD~135344B was fitted with a 2D Gaussian profile which yielded a typical FWHM of 41\,mas. More details on the observations and conditions are provided in Table~\ref{tab:observations}.

\subsection{Rapid Eye Mount (REM) visible and near-infrared photometry}\label{sec:rem}

HD~135344B was observed with the REM at La Silla, Chile, in June 2016. The visible camera, ROSS2, is a simultaneous multi-channel imaging camera which delivers the $g'r'i'z'$ bands onto four quadrants of the same CCD detector. A dichroic enables simultaneous observations with the infrared camera, REMIR, with a similar field of view of $10\arcmin \times 10\arcmin$. Observations were executed with 3\,s exposures in the $g'r'i'z'$ bands and 1\,s exposures in the $JHK$ bands. For the $JHK$ photometry, a standard five-position dither pattern was used and additional sky frames were obtained. Data was acquired during a total of twelve nights but the data of two nights were rejected due to thick clouds.

The photometry of HD~135344B was measured differentially with respect to HD~135344A (SAO~206463), an A0V star with a separation of $21\arcsec$ from HD~135344B, which appears to be photometrically stable \citep{sitko2012}. Differential photometry allowed us to measure with high precision the absolute fluxes of HD~135344B also in variable conditions or with the presence of thin clouds. The $JHK$ magnitudes of HD~135344A were retrieved from the 2MASS catalog \citep{skrutskie2006} while the $g'r'i'z'$ magnitudes were calculated through a transformation of the $BVRI$ magnitudes with the relations from \citet{jordi2006}. The $BVRI$ magnitudes of HD~135344A were obtained from \citet{sitko2012}: $B=7.879\pm0.003$\,mag, $V=7.756\pm0.003$\,mag, $R=7.708\pm0.006$\,mag, and $I=7.662\pm0.004$\,mag.

\subsection{PIONIER interferometry}\label{sec:pionier}

We retrieved all the available archival near-infrared interferometric data of HD~135344B from the PIONIER instrument \citep{lebouquin2011} at the Very Large Telescope Interferometer (VLTI). The data were taken during multiple epochs from 2011 to 2013\footnote{UT dates: 2011 April 27 and 29, 2011 May 26 and 27, 2011 June 3, 2011 August 7 and 8, 2012 March 6, 28, 29 and 30, 2012 April 27 and 2013 May 15.}. The instrument recombines the four auxiliary telescopes which were positioned in the short (A1-B2-C1-D0), intermediate (D0-G1-H0-I1), and long (A1-G1-I1-K0) baseline configurations. The projected baseline, $B$, ranged from 7 to 135\,m, enabling a maximum angular resolution of $\lambda/2B = 1.2$\,mas across the seven spectrally dispersed channels in the $H$ band ($\mathcal{R}\simeq40$). Each observation of HD~134453B was preceded and followed by an observation of a calibration star to characterize the instrumental and atmospheric contribution to the visibilities and closure phases (i.e., the transfer function). Calibration stars were identified with the \texttt{SearchCal} tool \citep{bonneau2006,bonneau2011}. The data were reduced with the \texttt{pndrs} package, described in detail by \citet{lebouquin2011}. The 27 calibrated OIFITS files \citep{pauls2005} are available in the \texttt{Optical interferometry DataBase (OiDB)} \citep{haubois2016}.

The observing conditions were best during the two epochs in 2011 April with a coherence time of $\sim$5\,ms and a seeing of 0\ffarcs5--0\ffarcs7, while the conditions were significantly poorer when the other data sets were taken. We rejected low S/N measurements by selecting only the data points where the error estimates were in the first quartile of the total distribution of errors. The dominating factor in the error estimate is the stability of the transfer function during the night which is determined by the temporal scatter of the calibrator visibilities. For each data set, the quality assessment was done independently for the six visibilities, four closure phases, and seven spectral channels. The selected measurements were retrieved from 11 out of 13 epochs, with 2011 May 26 and 2012 April 27 excluded.

\section{Results}\label{sec:results}

\subsection{Multi-epoch polarized light imagery}\label{sec:imagery}

\begin{figure*}
\centering
\includegraphics[height=19.33cm]{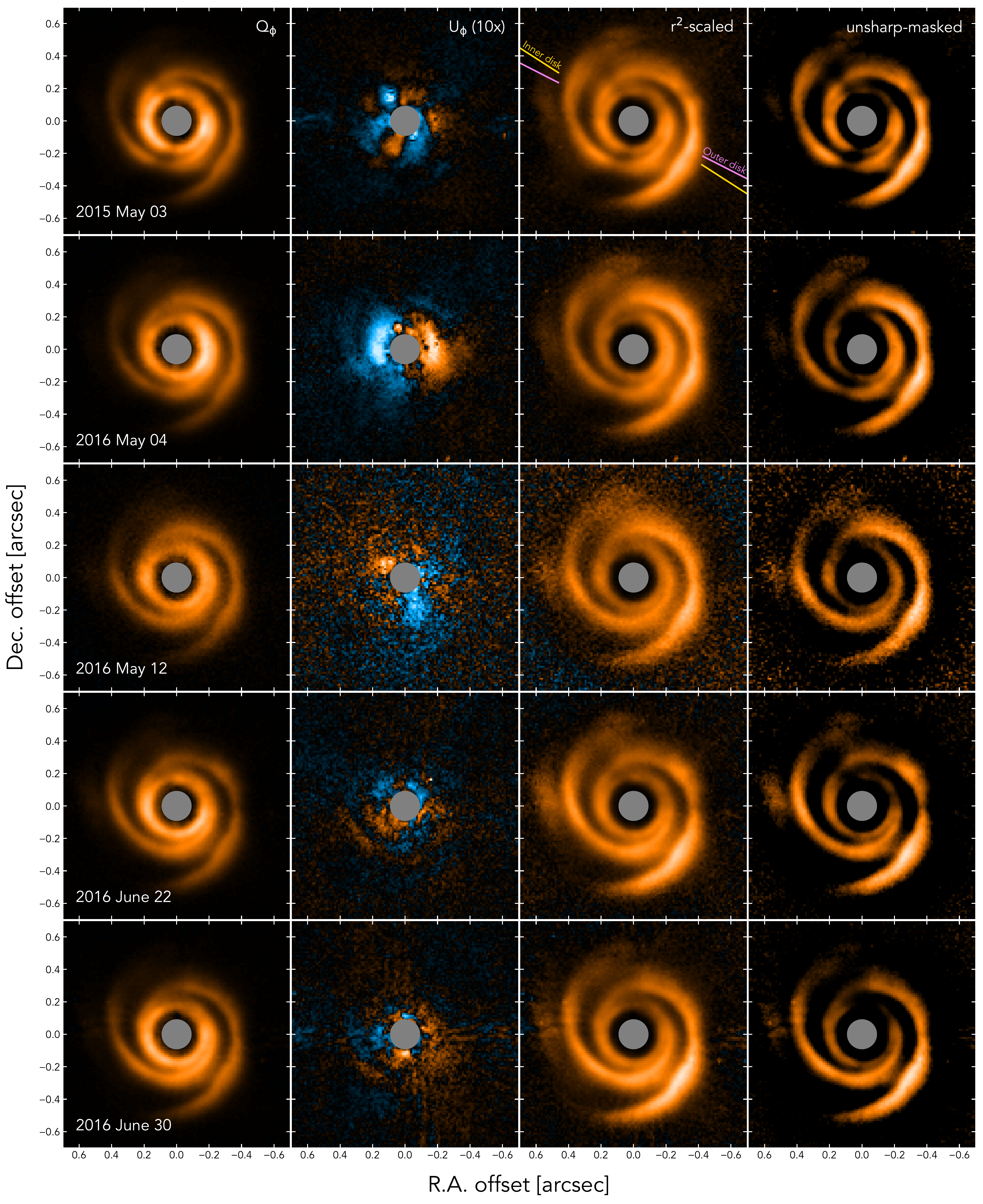}
\caption{Multi-epoch polarized scattered light images of HD~135344B in the $J$ band. The columns show from left to right the unscaled $Q_\phi$ images, unscaled $U_\phi$ images, $r^2$-scaled $Q_\phi$ images, and an unsharp-masked version of the $r^2$-scaled $Q_\phi$ images. The field of view of each image is $1\ffarcs4 \times 1\ffarcs4$ with north and east in upward and leftward direction, respectively. The surface brightness of the images has been normalized to the integrated $Q_\phi$ flux (see main text for details). The dynamical range of the color stretch is fixed in each column except for the unsharp-masked images. The dynamical range of the $U_\phi$ images is a factor 10 smaller than the partner $Q_\phi$ images. Orange corresponds to positive values, blue to negative values, and black is the zero point. The extent of the coronagraph has been masked out. The major axis position angle of the outer disk, ${\rm PA} = 63\degr$ \citep[purple line;][]{vandermarel2015}, and the inner disk, ${\rm PA} = 57\ffdeg3 \pm 5\ffdeg7$ (yellow line; see Section~\ref{sec:fit}), are shown in the top row.\label{fig:gallery}}
\end{figure*}

Scattered light images are displayed in chronological order (top to bottom) in Figure~\ref{fig:gallery}. Besides the newly obtained data sets from 2016, we also show the 2015 $J$-band imagery from \citet{stolker2016}. The first column shows the unscaled $Q_\phi$ images, defined such that positive values correspond to azimuthally polarized flux. The images are normalized to the disk-integrated $Q_\phi$ flux, measured with an annulus aperture between 0\ffarcs1 and 1\ffarcs0, and shown with identical dynamical range. The second column shows the corresponding $U_\phi$ images, containing flux with a $\pm45\degr$ rotational offset of the direction of polarization with respect to the $Q_\phi$ flux. The third column contains the $Q_\phi$ images with a stellar irradiation correction applied (i.e., $r^2$-scaling). The fourth column shows unsharp-masked images that were obtained by smoothing the $r^2$-scaled $Q_\phi$ images with a Gaussian kernel ($\sigma=200$\,mas) and subtracting the smoothed images from the original $r^2$-scaled images. This procedure enhances the contrast of small scale features by removing low spatial frequencies. The dynamical range of the unsharp-masked images is limited to positive values and for each image separately normalized to the peak intensity.

The SW direction (${\rm PA} = 180\degr$--$270\degr$) of the disk, located around the major axis (see Figure~\ref{fig:gallery}), appears relatively bright in all $r^2$-scaled images. In contrast to the shadowing variations, the origin of that brightness enhancement is presumably intrinsic since the bright wedge in scattered light coincides with the (sub)millimeter emission peak of the crescent-shaped dust continuum \citep{perez2014,stolker2016}. An enhancement of the surface density and/or midplane temperature will elevate the height of the scattering surface, therefore, a larger geometrical cross section of the disk surface is irradiated which increases the scattered light flux. In this work, we focus on local brightness variations between the five epochs. We refer the reader to \citet{muto2012}, \citet{garufi2013}, \citet{stolker2016} for a detailed analysis and discussion of the scattered light detection of the spiral arms and cavity edge.

The $U_\phi$ images in Figure~\ref{fig:gallery} contain contributions from noise, residuals of instrumental polarization, and possibly multiple scattered light from the disk. The residual signal could not be removed with the minimization steps explained in Section~\ref{sec:irdis} but a more detailed analysis is required to determine if the remaining signal is an instrumental artifact. We distinguish between two different type of signals in the $U_\phi$ images which appear to be related to the total integration time and therefore the S/N. First, the relative contribution of noise is particularly well visible at separations $\gtrsim 300$\,mas from the star in the last three epochs for which the total integration time was shortest while the relative contribution is lower in the first two epochs. Second, the $U_\phi$ images show an enhanced signal within 300\,mas of which the relative strength is larger in the first two epochs. The inner signal reaches only mildly above the background noise level in the remaining epochs. The inner $U_\phi$ signal appears to be variable between epochs revealing a variety of brightness patterns. For example, the first epoch shows a complex pattern of multiple positive and negative lobes whereas the second epoch displays an anti-symmetric signal which is bisected in a positive and negative side. The relative strength of the $U_\phi$ flux with respect to the $Q_\phi$ flux is quantified in Section~\ref{sec:photometry}.

Although we deem it likely that the remaining $U_\phi$ signal is an uncorrected instrumental artifact, we speculate that the increasing strength of the innermost $U_\phi$ signal with increasing S/N could be a result of multiple scattered light from the cavity edge. A minor fraction of the scattered light will be non-azimuthally polarized because of the small inclination of the outer disk \citep[$16\degr$--$20\degr$;][]{vandermarel2015,vandermarel2016}. Furthermore, a small fraction of the stellar light will scatter in the extended inner disk atmosphere (see Section~\ref{sec:radiative_transfer}) before it scatters from the outer disk. The $U_\phi$ signal from those photons will get modulated by the temporal variations in the inner disk such that each epoch may show a different $U_\phi$ image. However, more detailed analysis and modeling is required to determine if the $U_\phi$ signal is real and if the temporal variations could be caused by the inner disk.

\begin{figure}
\centering
\includegraphics[width=6.85cm]{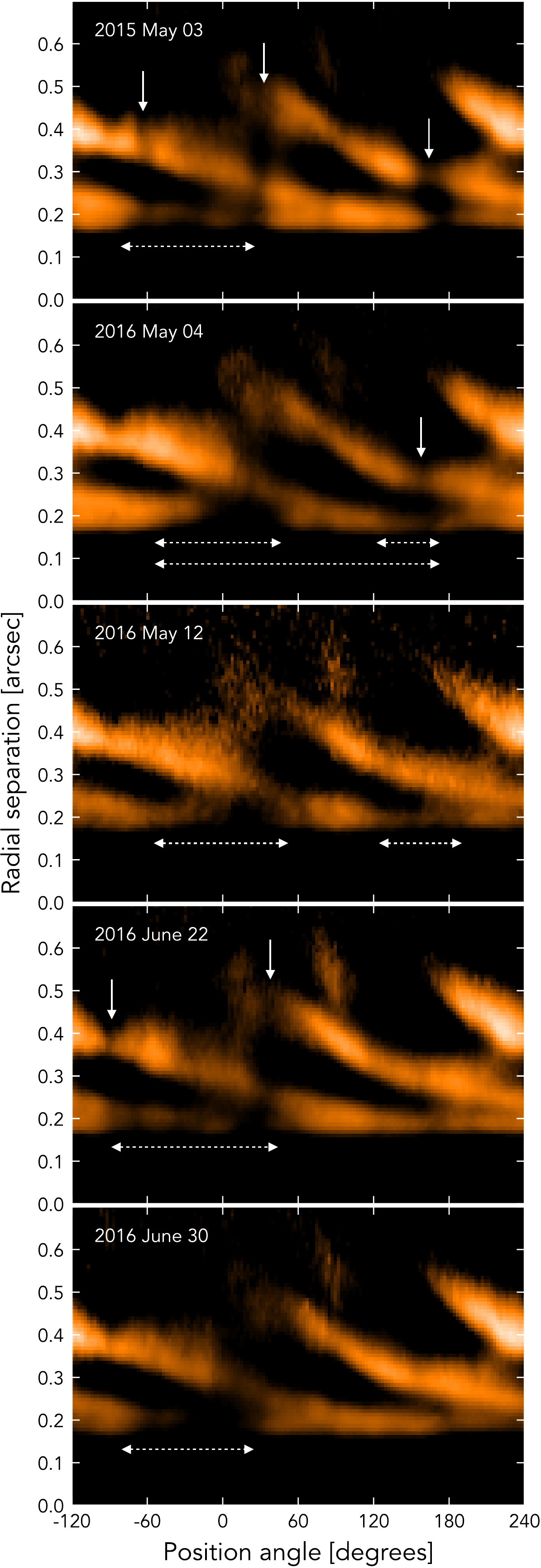}
\caption{Polar projections of the $r^2$-scaled, unsharp-masked $Q_\phi$ images shown in chronological order (top to bottom). North corresponds to ${\rm PA} = 0\degr$ and positive position angles are measured east from north. Localized and broad shadow features are indicated with solid and dashed arrows, respectively.\label{fig:polar}}
\end{figure}

\subsection{Asymmetric Q$_\phi$ brightness variations}\label{sec:brightness_variations}

A comparison of the $r^2$-scaled $Q_\phi$ images in Figure~\ref{fig:gallery} shows epoch-to-epoch brightness variations. Azimuthal brightness minima are visible in all images with variations in their location, shape, and strength. The locality of the brightness minima (e.g., the shadow lane at ${\rm PA} = 169\degr$ on 2015 May 03) points toward a shadowing effect, likely caused by dust in the (unresolved) inner disk \citep{stolker2016}. Furthermore, the brightness variations occur on a timescale similar to the dynamical timescale of the inner disk (the finest temporal resolution is 8\,days) and the variability timescale of the near-infrared photometry. Minor brightness variations in the unscaled $Q_\phi$ images are also visible in the region between the cavity edge and the coronagraph, possibly related to the flow of gas and small dust grains from the outer disk.

Azimuthal brightness variations, and their changes between epochs, are more evidently revealed with polar projections of the scattered light images which are displayed in Figure~\ref{fig:polar}. For clarity, we choose the unsharp-masked images for the identification of the shadow features in the polar projections. We caution that applying an unsharp mask may introduce a bias in the identification of brightness variations. However, the shadow features, that we will discuss below and are marked with arrows in Figure~\ref{fig:polar}, are also visible in the regular $r^2$-scaled $Q_\phi$ images, but the contrast between shadowed and non-shadowed regions is smaller.

Asymmetric illumination/shadowing variations are visible in the scattered light imagery of all five epochs. Here we list the main characteristics of the shadow features, in consonance with the locations that are pointed out in Figure~\ref{fig:polar}:
\begin{enumerate}
  \item Epoch 1, 2015 May 03 - Three narrow shadow lanes are present at position angles of $34\degr$, $169\degr$, and $304\degr$, and an azimuthally broader dimming is visible in N-NW direction which is bound by two of the shadow lanes \citep{stolker2016}. 
  \item Epoch 2, 2016 May 04 - The eastern half of the disk is mildly shadowed, approximately in the position angle range of $10\degr$--$170\degr$. Deeper shadows are superimposed near the edges of the global shadow. The deepening is particularly well visible in the north, extending at the cavity edge from ${\rm PA} = -50\degr$ to ${\rm PA} = 50\degr$. Radially outward, the location of the shadow shows an azimuthal gradient. There is a hint of a localized shadow lane at ${\rm PA} = 170\degr$, approximately colocated with the southern shadow lane in the first epoch.
  \item Epoch 3, 2016 May 12 - The bisection of the brightness distribution from the second epoch seems to have disappeared although the poor observing conditions and the short total integration time (see Table~\ref{tab:observations}) make the identification of the shadows challenging. The broad, northern shadow from the previous epoch is still present. There is also a hint of the broad, southern shadow whereas the narrow southern shadow has disappeared.
  \item Epoch 4, 2016 June 22 - A broad shadow is present between ${\rm PA} = -90\degr$ and ${\rm PA} = 30\degr$ upon which finer shadow variations are superimposed, including narrow shadow lanes at the boundary with the non-shadowed region, similar to the northern shadow features in the first epoch. The shadow lane at ${\rm PA} = 30\degr$ possibly coincides with the location of the NE shadow lane detected in first epoch.
  \item Epoch 5, 2016 June 30 - The cavity edge is shadowed between ${\rm PA} = -45\degr$ and ${\rm PA} = 40\degr$ while shadowing of the exterior spiral arm only occurs from ${\rm PA} = 0\degr$ onwards. The radial extent of the broad shadow increases with increasing position angle, similar to the broad shadows in the first and second epoch. The shadow covers the full radial extent of the disk between ${\rm PA} = 0\degr$ and ${\rm PA} = 40\degr$, similar to the shadow feature at the same location in the second and third epoch. The narrow shadow lanes from the fourth epoch seem to have disappeared.
\end{enumerate}

\begin{figure*}
\centering
\includegraphics[height=9cm]{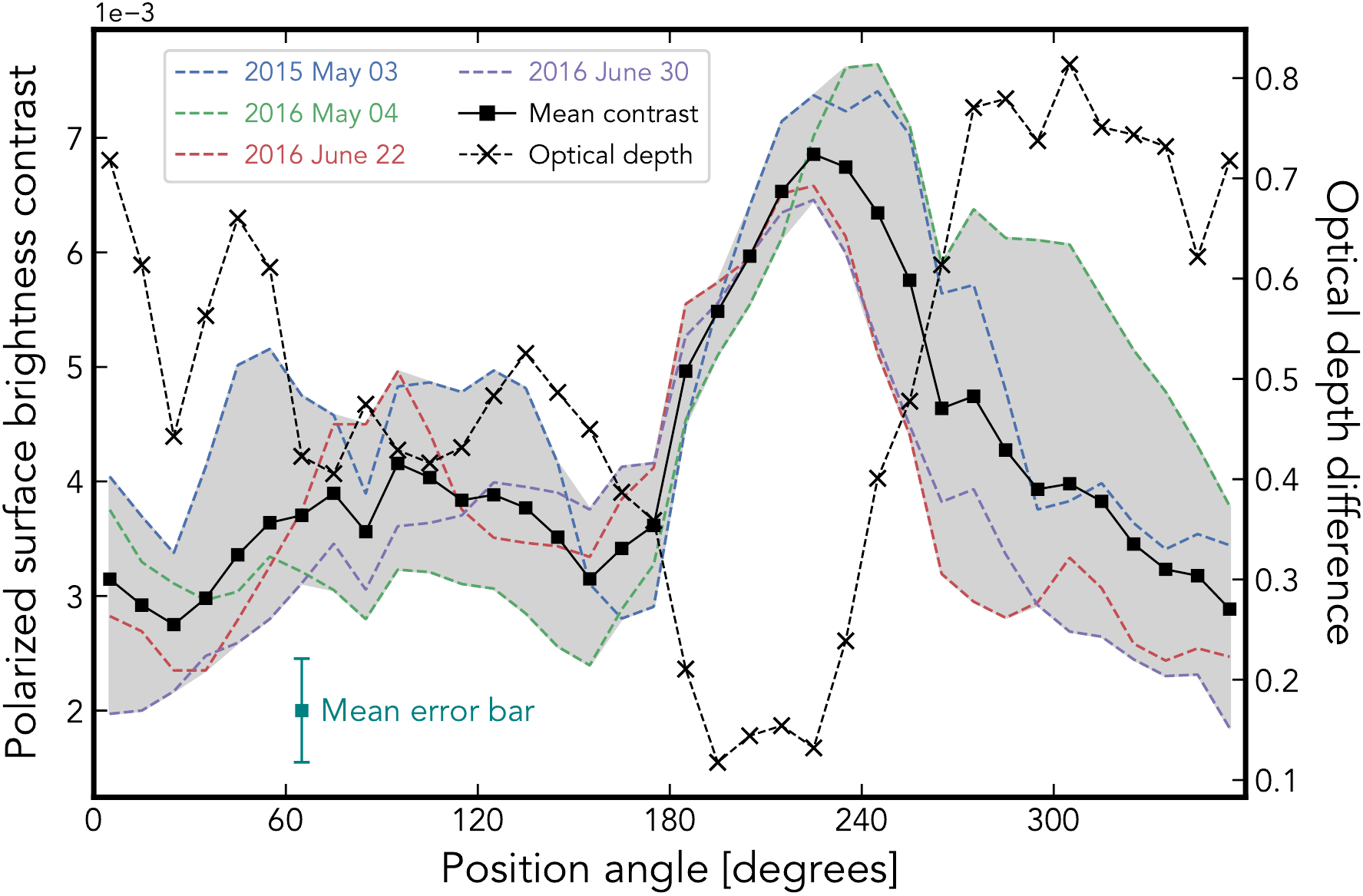}
\caption{Polarized brightness of the disk normalized to the total Stokes~$I$ flux. The plot shows the contrast from four of the epochs (colored dashed lines), the mean contrast (black solid line), and the relative optical depth variation between the minimum and maximum contrast (black dashed line, right $y$-axis). The gray shaded area covers the total variation of the contrast between the epochs. The mean error on the contrast, across all epochs and position angles, is shown on the bottom of the figure (see main text for details). The image from 2016 May 12 has been excluded since it was affected by the poor observing conditions.\label{fig:brightness}}
\end{figure*}

The brightness depth of the shadow lanes appears typically larger than the broader shadows (e.g., epoch 1 in Figure~\ref{fig:gallery}) indicating larger density enhancements in the inner disk. Indeed, there is no evident radial dependence in the depth of the shadow lanes even though the height of the outer disk increases with radius. The brightness depth of the broad shadows is typically deepest at the cavity edge but weakens toward larger radii (e.g., epoch 4 in Figure~\ref{fig:polar}), presumably an effect of the increasing outer disk height. The broad shadow in N-NW direction (${\rm PA} \simeq -80\degr$--\,$30\degr$) of all epochs shows an azimuthal gradient which is possibly a result of the asymmetric spiral arm perturbation of the disk surface.

The absence of strong azimuthal gradients, caused by a light travel time effect, in the location of the narrow shadows provides a lower limit on the radius from where the shadows are cast. For example, the position angle of a shadow will change by $10\degr$ between the inner disk and 80\,au (500\,mas) if the responsible dust clump is located at 0.15\,au \citep[see also][]{kama2016}. Arguably, some of the shadow lanes show very minor tilts in Figure~\ref{fig:polar} but the precision and angular resolution of the observations challenge the identification of light travel time effects. The best candidate is the shadow lane at ${\rm PA} = 169\degr$ in the 2015 epoch. \citet{stolker2016} speculate that its azimuthal tilt could be caused by orbital motion in the inner disk from which an orbital radius of 0.06\,au (at 140\,pc) was estimated. For the other shadow lanes, we may conclude that the dust clumps responsible are presumably located at distances $\gtrsim$0.15\,au.

A quantification of the azimuthal brightness variations is shown in Figure~\ref{fig:brightness}. The mean $Q_\phi$ flux is measured in position angle bins of $10\degr$ wide across a radial separation of 0\ffarcs1--0\ffarcs7 and divided by the angular area of a pixel. The polarized surface brightness (in counts\,s$^{-1}$\,arcsec$^{-2}$) is normalized to the total Stokes~$I$ flux (in counts\,s$^{-1}$) which is measured with a circular aperture on the unsaturated, non-coronagraphic flux frames after a correction for the integration time and response of the neutral density filter. The optimal aperture size (1\ffarcs5) was determined by measuring the photometric flux with a large range of aperture sizes (up to 3\ffarcs0) from which it was established that the total encompassed flux flattened for apertures larger than $\sim$$1\ffarcs5$. The mean error bar is calculated from the standard error on the individual contrast points.

\begin{figure*}
\centering
\includegraphics[height=7.5cm]{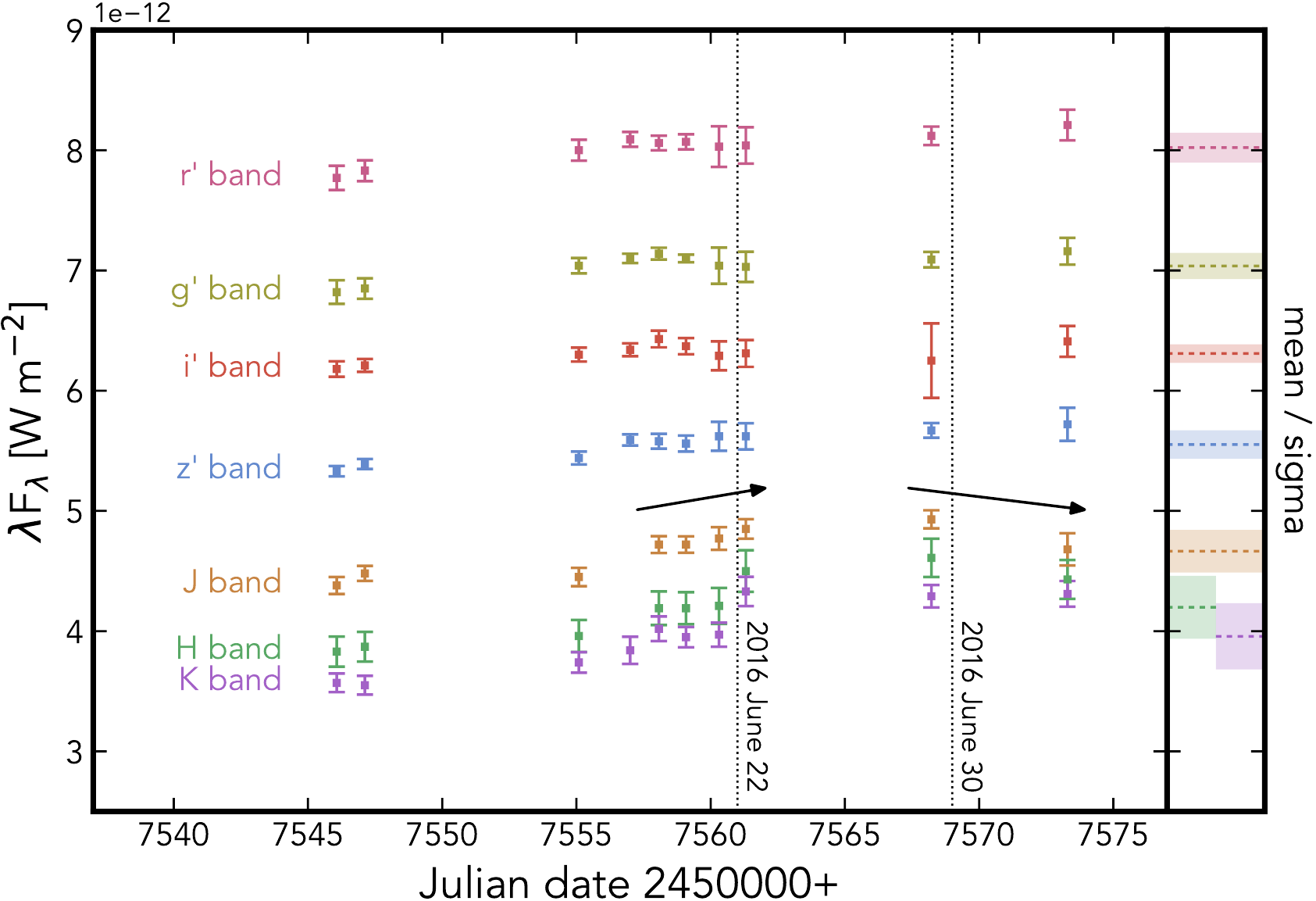}
\caption{Visible and near-infrared photometry obtained with the Rapid Eye Mount during 10 nights in 2016 June. The mean and standard deviation of the fluxes are provided by the horizontally dashed lines and shaded regions, respectively. Photometric monitoring overlapped with the two most recent SPHERE epochs of which the UT dates are indicated with vertically dotted lines. The arrows point in the direction of an increasing and decreasing near-infrared variation.\label{fig:photometry}}
\end{figure*}

The polarized surface brightness contrast in Figure~\ref{fig:brightness} shows typical values in the range of 2--6$\times 10^{-3}$, except in the SW direction where the contrast goes up to $8 \times 10^{-3}$ at ${\rm PA}\simeq240\degr$. The integrated disk brightness consists mainly of signal from the cavity edge and the spiral arms (see unscaled $Q_\phi$ images in Figure~\ref{fig:gallery}) which are both asymmetrically shadowed resulting in dimming variations of 20--45\%. The contrast variation is minimal (10\%) in S-SW direction where the large-scale density and/or scale height enhancement affects the disk surface. The opening angle of the scattering surface is larger in that direction such that shadowing by the inner disk requires dust to be located at higher altitude above the midplane. The contrast variations provide an upper limit on the local changes in optical depth through the inner disk atmosphere. The stellar radiation that is transmitted through the atmosphere will be attenuated by a factor $e^{-\tau}$, therefore, the relative change in optical depth, $\Delta\tau$, can be calculated from the minimum and maximum contrast (see Figure~\ref{fig:brightness}). However, this only provides an upper limit on the optical depth variations because (i) the total flux does also affect the contrast and is correlated with the shadowing of the outer disk (see Section~\ref{sec:radiative_transfer}), (ii) part of the thermal emission from the inner disk may illuminate the outer disk without being affected by any local obscurations, and (iii) PSF smearing will lower the brightness contrast between shadowed and non-shadowed regions.

\subsection{Photometry and scattered light contrast}\label{sec:photometry}

Multi-epoch photometry in the $g'r'i'z'$ and $JHK$ bands are displayed in Figure~\ref{fig:photometry}, covering ten nights between 2016 June 07 and 2016 July 04 with nearly a daily sampling from June 16 till June 22. The error bars reflect the uncertainty on both the science and calibration star. The early June photometry in the $g'r'i'z'$ bands starts with a minor increase of $3\%$, remains approximately constant in mid-June, and increases with $2\%$ by the end of June.

The temporal course of the $JHK$ photometry appears more irregular with variations up to 10\% with respect to the mean (horizontally dashed lines in Figure~\ref{fig:photometry}) in the $JHK$ bands. During the first half of June 2016, the fluxes increased with approximately 6\%, 9\%, and 10\% in the $J$, $H$, and $K$ bands, respectively, following the trend of the $g'r'i'z'$ fluxes but with a larger fractional increase. In mid-June, also the $JHK$ fluxes remained approximately constant but they increased further from June 22 onwards, in contrast to the $g'r'i'z'$ fluxes. The final epoch shows a decline in the $J$ and $H$ bands while the $K$ band photometry remained constant. Although the $JHK$ variability in the first part of June seems correlated with the $g'r'i'z'$ photometry, in the second half no correlation is apparent, that is, the $JHK$ fluxes increased up to 10\% with respect to the mean while the $g'r'i'z'$ photometry remained constant. The total temporal coverage of the photometry is too short to reveal any trends and the sampling is too sparse to resolve possible variations on timescales less than one day.

\begin{figure}
\centering
\includegraphics[width=\linewidth]{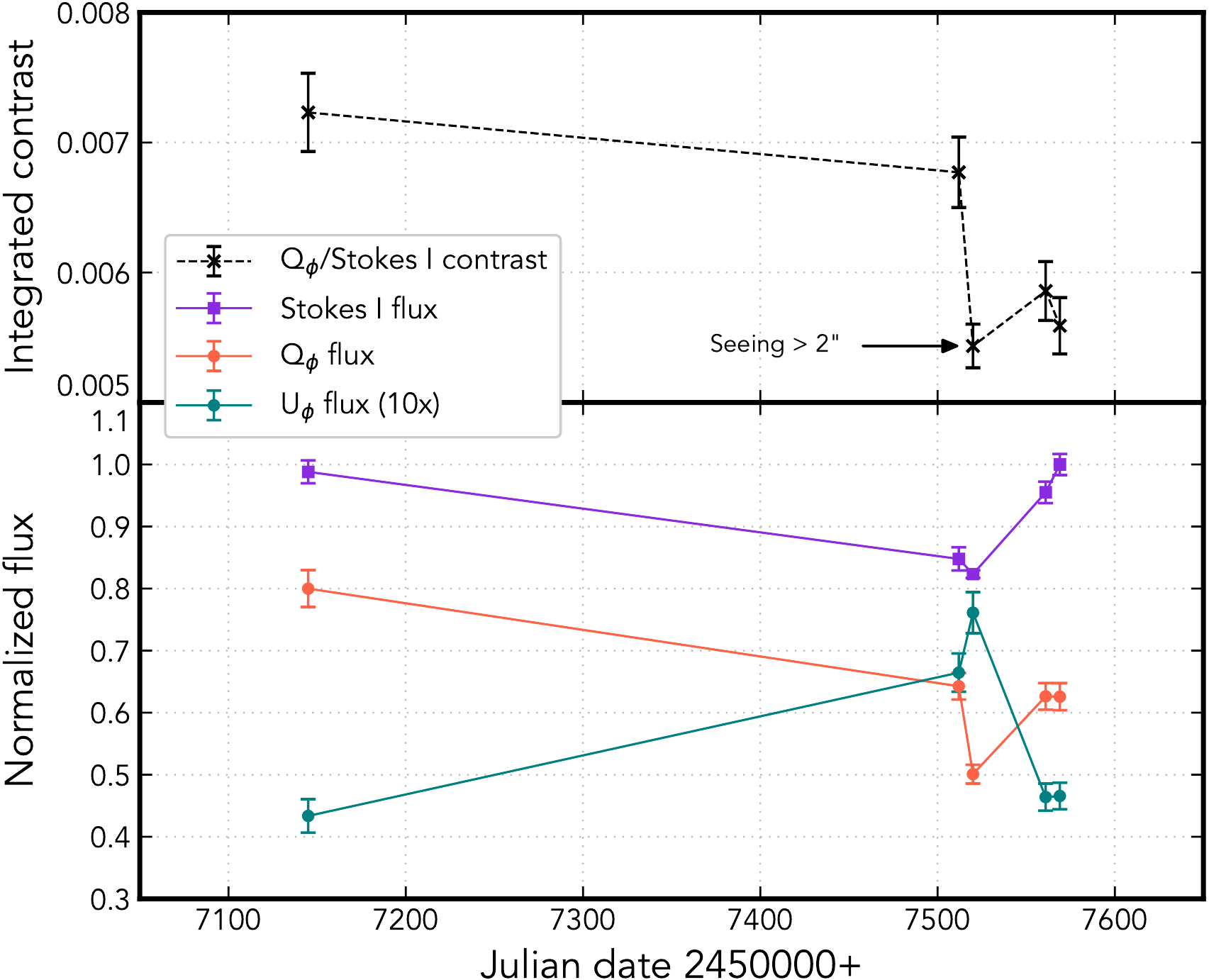}
\caption{\emph{Top}: integrated polarized scattered light contrast (black crosses) of the five $J$-band SPHERE epochs. \emph{Bottom}: the integrated Stokes~$I$ flux (purple squares) and the disk-integrated $Q_\phi$ flux (red circles), shown in arbitrary normalized units. The disk-integrated $U_\phi$ flux (green circles) is computed from the absolute pixel values and shown relative to the $Q_\phi$ flux with a factor 10 enhancement. The uncertainties are given at a $5\sigma$ level.\label{fig:contrast}}
\end{figure}

In addition to the absolute REM photometry, we measured the relative disk photometry of the SPHERE data which is presented in Figure~\ref{fig:contrast}. The disk-integrated polarized flux was determined from the $Q_\phi$ and $U_\phi$ images with an annulus aperture (0\ffarcs1--1\ffarcs0) centered on the star. The Stokes~$I$ flux was measured with a circular aperture (1\ffarcs5 radius) from the dedicated flux frames. Absolute pixel values were used for the $U_\phi$ photometry. The photometric contrast (top panel in Figure~\ref{fig:contrast}) is calculated as the ratio of the $Q_\phi$ and Stokes~$I$ flux after correcting for the difference in integration time and the response of the neutral density filter. Relative photometry allows for an epoch-to-epoch analysis without requiring an absolute flux calibration, assuming that both the coronagraphic sequence and the total flux data were obtained during similar observing conditions. The $Q_\phi$ and $U_\phi$ surface brightness errors are computed as the standard deviation within an aperture centered on each pixel with a radius of 62\,mas (i.e., 1.5 resolution elements) and propagated accordingly to an integrated error (3\%--5\%). The uncertainty on the total flux (1\%--2\%) is computed in a background-limited region as the standard error of the sum, $\sigma\sqrt{N_{\rm pix}}$, with an annulus aperture equal in size to the aperture used for the Stokes~$I$ photometry.

The integrated contrast in Figure~\ref{fig:contrast} varies between $5.4 \times 10^{-3}$ and $7.2 \times 10^{-3}$. The photometry is not calibrated so only relative variations of $Q_\phi/{\rm Stokes}~I$ and $U_\phi/Q_\phi$ are meaningful. The second epoch shows a consistent decrease of both the $Q_\phi$ and Stokes~$I$ flux with respect to the first epoch. In the third epoch, the contrast decreased by 20\% possibly due to the poor observing conditions (see Section~\ref{sec:irdis}), particularly affecting the $Q_\phi$ photometry. During the last two epochs, the relative increase of the total flux is large compared to the $Q_\phi$ flux, resulting in relatively low contrast. The increase of the total flux during the last epochs seems consistent with the REM photometry (see June 22 and 30 in Figure~\ref{fig:photometry}). The relative $U_\phi$ photometry shows an increase in the second epoch due to the residual within 200\,mas (see Section~\ref{sec:imagery}), possibly caused by multiple scattered light from the inner and/or outer disk, while the relative $U_\phi$ photometry in the third epoch is larger due to the enhanced noise residual.

\subsection{Parametric model fitting of the visibilities}\label{sec:fit}

\begin{figure*}
\centering
\includegraphics[width=16cm]{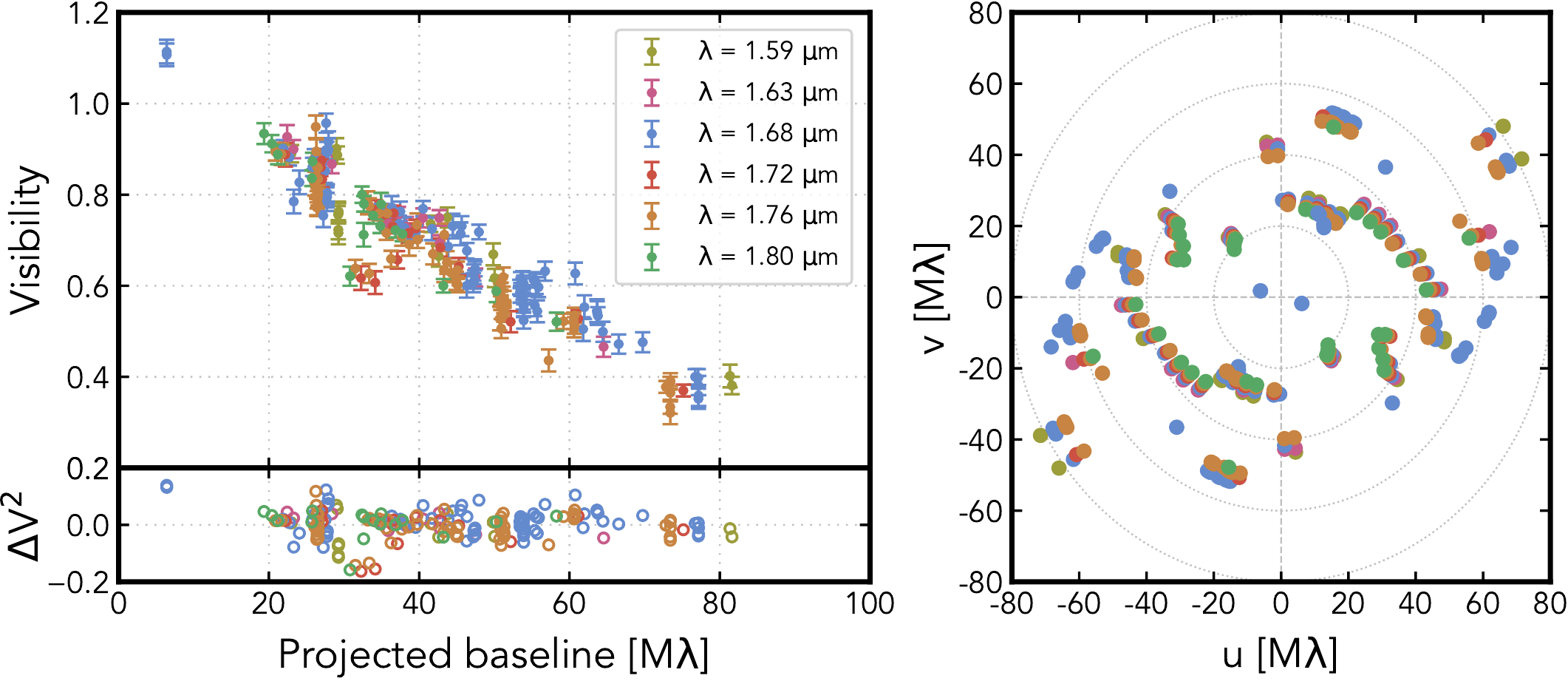}
\caption{\emph{Left:} squared visibilities ($V^2$) of in the VLTI/PIONIER $H$-band channels with $\mathcal{R}\simeq40$ (top) and the fitting residuals of the ellipsoidal brightness model (bottom). \emph{Right}: coverage of the $(u,v)$-plane, shown with the same color coding as the visibilities.\label{fig:pionier}}
\end{figure*}

The normalized, squared visibilities, $V^2$, across the spectrally dispersed $H$-band channels of the multi-epoch PIONIER observations are displayed in the left panel of Figure~\ref{fig:pionier}. The visibilities decrease continuously with increasing spatial frequency (i.e., \mbox{$B/\lambda$}), indicating that the region from which the $H$-band flux originates is resolved. At the longest baselines, there appears no turnover point to an asymptotic value of the stellar flux so the circumstellar emission is not over-resolved. The closure phases, $\Delta\phi$, are consistent with zero within the error bars ($\left|\Delta\phi\right|\lesssim 3\degr$), therefore, the brightness distribution of the inner disk is point symmetric within the uncertainties and at the spatial resolution probed by the observations. There is no significant dispersion visible, for visibility points obtained with baselines of similar lengths but different position angles, which implies that the inclination of the inner disk is small. The diagonal scatter of the visibilities is an effect of chromatic dispersion \citep{lazareff2017}. Coverage of the $(u,v)$-plane is shown in the right panel of Figure~\ref{fig:pionier}.

The orientation and characteristic radius of the inner disk $H$-band emission is determined by fitting, in Fourier space, a parametric model to the visibilities, following the procedure described in detail by \citet{lazareff2017}. This allowed us to apply a $\chi^2$ minimization and to assign formal error bars to the inferred parameter values. We parameterize the $H$-band emission with an elliptical brightness distribution that is radially parameterized by a weighted combination of a Gaussian and pseudo-Lorentzian profile. The inner rim is not fully resolved by the longest baselines which justifies an ellipsoidal distribution instead of a broadened ring.

The best-fit model ($\chi^2=1.10$) corresponds to an inclination and major axis position angle of $18\ffdeg2^{+3.4}_{-4.1}$ and $57\ffdeg3 \pm 6\ffdeg3$, respectively. The values are, within the uncertainties, very similar to those of the outer disk (see discussion in Section~\ref{sec:interferometry_constraints}). The best-fit position angle is shown in Figure~\ref{fig:gallery} in comparison with the outer disk value from \citet{vandermarel2015}. The half-flux semi-major axis is $0.71 \pm 0.03$\,mas ($=0.11$\,au) which implies that a significant fraction of the $H$-band emission originates from within the silicate sublimation radius \citep[$R_{\rm sub}=0.2$\,au;][]{carmona2014}. A detailed overview of the fitting results is provided in Appendix~\ref{sec:appendix} where the best-fit values of all parameters are listed with their dependence on the cutoff level of the selection criterion for the $(u,v)$ points.

We caution that the visibilities were combined from multiple epochs while the inner disk is variable on a timescale of days or less (see Section~\ref{sec:photometry}). The visibilities in Figure~\ref{fig:pionier} depend on the absolute $H$-band flux and the relative contributions of the star and disk. Therefore, an additional uncertainty has been introduced by combining the visibilities from multiple epochs while the absolute $H$-band flux is variable and not measured at the nights of the observations. Also, we made the assumption that the orientation of the inner disk did not change due to precession between those epochs.

\section{Discussion}\label{sec:discussion}

In Section~\ref{sec:misalignment} we will discuss the available constraints on the orientation of the inner and outer disk, as well as the presence of the broad, quasi-stationary shadow. In Section~\ref{sec:origin}, we will provide an observational perspective on the variability and we will discuss several processes that may affect the inner disk dynamics. In Section~\ref{sec:radiative_transfer}, we will estimate the extent of the inner disk variations by quantifying the correlation between the thermal emission and the scattered light flux with a grid of radiative transfer models.

\subsection{Constraints on the inner disk (mis)alignment}\label{sec:misalignment}

\subsubsection{Near-infrared and submillimeter interferometry}\label{sec:interferometry_constraints}

The relative orientation of the inner and outer disk is determined by their inclination with respect to the plane of the sky and their position angle, with the misalignment defined by the angle between the normal vectors of the two midplane orientations. The outer disk's orientation of HD~135344B has been determined in several studies \citep[see][for an overview]{carmona2014}. Here we list those values that were derived from spatially resolved (sub)millimeter CO observations: $i=11\ffdeg5\pm0\ffdeg5$ and ${\rm PA}=64\degr\pm2\degr$ \citep{lyo2011}, $i=20\degr$ and ${\rm PA}=63\degr$ \citep{vandermarel2015}, $i=16\degr$ and ${\rm PA}=63\degr$ \citep{vandermarel2016}. In Section~\ref{sec:fit}, we found that the inner disk inclination and position angle are $i = 18\ffdeg2^{+3.4}_{-4.1}$ and ${\rm PA} = 57\ffdeg3 \pm 5\ffdeg7$, respectively. The values are consistent with the orientation of the outer disk within the uncertainties of the model fitting and the available values for the outer disk. This result is in contrast to several earlier studies which suggested a significant misalignment between the inner and outer disk \citep{fedele2008,grady2009,stolker2016}. However, we can not exclude a minor disk warp given the uncertainties on both the inner and outer disk orientation.

\begin{figure*}
\centering
\includegraphics[width=15cm]{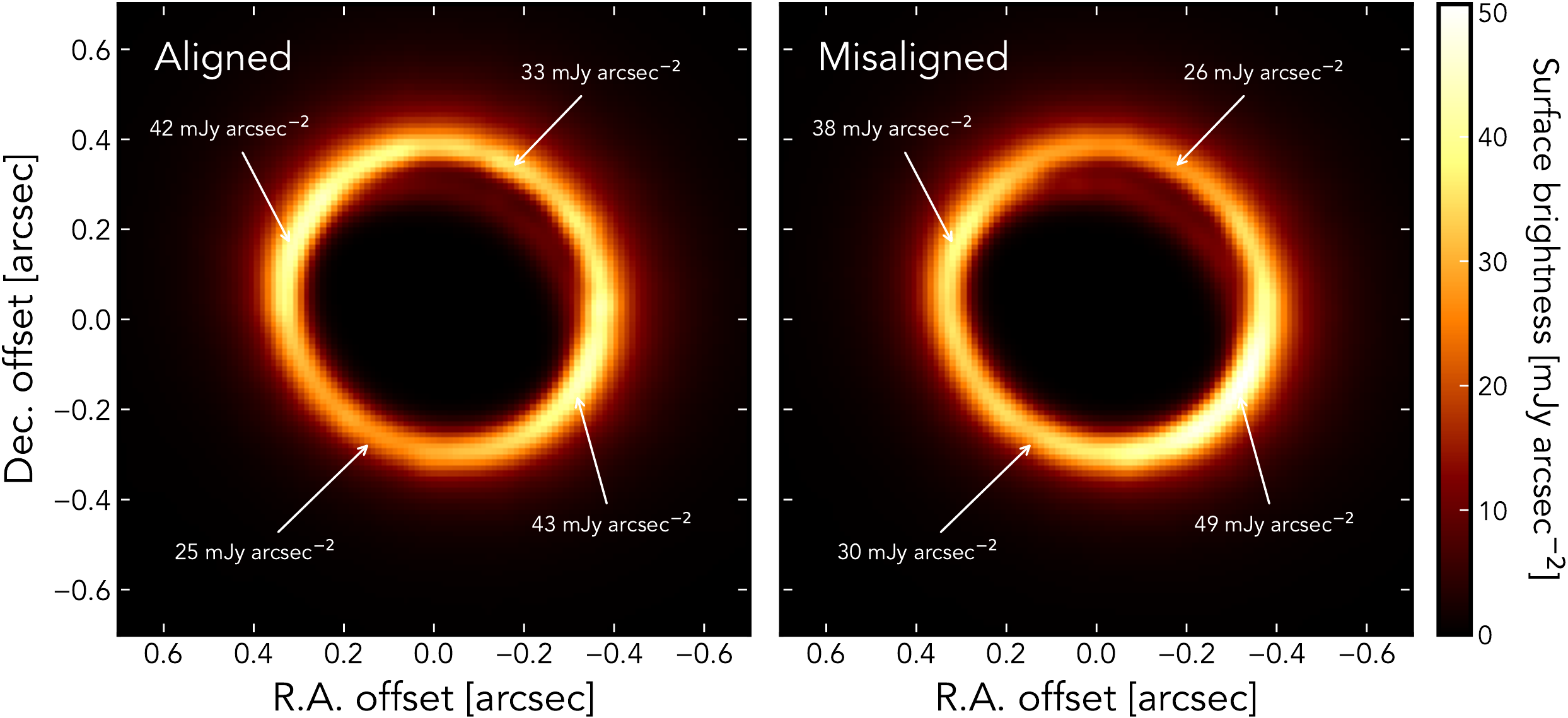}
\caption{Raytraced $Q_\phi$ images of the DIANA radiative transfer model of HD~135344B \citep{woitke2016}. \emph{Left}: the inner disk is aligned with the outer disk. \emph{Right}: the misalignment of the inner disk is $2\ffdeg6$, by adopting the best-fit inclination and position angle from Section~\ref{sec:fit}. Images have been convolved with a Gaussian kernel (${\rm FWHM} = 41$\,mas) to match the angular resolution of the SPHERE imagery. Surface brightness values are provided along the major and minor axis direction of the outer disk.\label{fig:warp}}
\end{figure*}

A possible misalignment of the two disk components relies also on the identification of the near and far side of both the inner and outer disk. Spatially resolved observations of CO gas show a redshifted and blueshifted velocity in SW and NE direction, respectively \citep[e.g.,][]{perez2014}. This implies that the near side of the outer disk is along SE direction of the minor axis if we assume that the spiral arms in scattered light follow a trailing motion. For the inner disk, there is no direct constraint on the near and far side at the angular resolution of the \mbox{PIONIER} observations. However, the misalignment will be $\sim$$38\degr$ if the near side of the inner disk is in NW direction, a scenario that can be excluded from the absence of two stationary shadow lanes similar to HD~142527 \citep[$\Delta\theta=70\degr$;][]{marino2015} and HD~100453 \citep[$\Delta\theta=72\degr$;][]{benisty2017}. This means that the near side of the inner disk is located in SE direction and a possible misalignment will be minor.

\subsubsection{Quasi-stationary shadowing by a minor disk warp?}\label{sec:warp}

The location and width of the shadows provide further constraints on the orientation of the inner disk with respect to the outer disk. Three scenarios remain possible: (i) the inner disk and outer disk are aligned such that shadowing occurs only through uplifting of dust in the inner disk atmosphere, (ii) a minor misalignment ($\Delta\theta\sim1\degr$--$2\degr$) is present which might not cast a stationary shadow but additional variations in the inner disk will cause preferential shadowing in the direction where elevation of the inner disk above the outer disk midplane is largest, (iii) an intermediate misalignment ($\Delta\theta\sim2\degr$--$10\degr$) is present which casts a broad, stationary shadow when the misalignment of the inner disk is similar to the opening angle of the scattering surface of the outer disk. An example of the third scenario is the broad shadow detected with \emph{HST} on the TW~Hya disk \citep[$\Delta\theta=8\degr$;][]{rosenfeld2012}.

The inclination of the inner disk could be either smaller or larger than the outer disk given the uncertainties on its orientation (see Section~\ref{sec:pionier}) which could result in a broad shadow in approximately NW or SE direction, respectively. Interestingly, a broad shadow seems present in all images in N-NW direction (see Figure~\ref{fig:polar}) although with small variations in its precise location, shape, and depth. This may imply that the inclination of the inner disk is slightly smaller (i.e., more face-on) than the outer disk. Variable fine structure is present in the N-NW shadow which requires additional optical depth variations through the atmosphere of the inner disk as will be discussed in more detail in Section~\ref{sec:origin}.

A broad dimming was also present in NW direction of the Subaru/HiCIAO $H$-band imagery by \citet{muto2012} and the VLT/NACO $H$ and $K_{\rm s}$-bands imagery by \citet{garufi2013}. This was interpreted in both studies as a depolarization effect which can occur when the disk is inclined because the polarization efficiency peaks around the major axis where scattering angles are close to $90\degr$ \citep[e.g.,][]{murakawa2010,min2012}. The inclination of the outer disk is relatively small so a strong depolarization effect might not be expected. We speculate that the broad dimming in the HiCIAO and NACO imagery might be the same quasi-stationary shadowing effect that is seen in the SPHERE imagery, possibly related to a minor disk warp.

To illustrate this scenario, we have adopted the \mbox{DIANA} \citep[Disc Analysis;][]{woitke2016} radiative transfer model of HD~135344B. The left image in Figure~\ref{fig:warp} displays the raytraced $Q_\phi$ image for a setup in which the inner disk is aligned with the outer disk \citep[$i=20\degr$, ${\rm PA}=63\degr$;][]{vandermarel2015}. A mild depolarization effect is seen in both directions of the minor axis but the effect is slightly stronger on the near side (SE) due to the flaring geometry of the disk surface \citep{min2012}. In the right image of Figure~\ref{fig:warp}, we changed the orientation of the inner disk to the best-fit values from Section~\ref{sec:fit}. The misalignment here is $2\ffdeg6$, that is, comparable to the disk warp seen in the debris disk around $\beta$~Pic \citep[$\Delta\theta=4\ffdeg6$;][]{heap2000}. The atmosphere of the inner disk casts a mild shadow on the outer disk in NW direction, reminiscent of the broad shadow on the cavity edge of the SPHERE images in Figures~\ref{fig:gallery} and \ref{fig:polar}. In opposite direction, the outer disk becomes more strongly irradiated which introduces an azimuthal brightness modulation along the cavity edge \citep[see also][]{rosenfeld2012}, although intertwined with the modulation by the polarization efficiency of the outer disk.

\subsection{Origin of the shadows and their variability}\label{sec:origin}

\subsubsection{Observational perspective and variability timescale}\label{sec:observational_timescales}

The most prominent azimuthal brightness variations were identified in Section~\ref{sec:brightness_variations} and we noticed that the shadows can be classified into two categories, (i) localized shadow lanes and (ii) broader shadows that are tens of degrees wide. The shadow variations are presumably caused by small density enhancements in the atmosphere of the inner disk with the optical depth variations being approximately smaller than unity (see Figure~\ref{fig:brightness}). Although a broad shadow is present in N-NW direction of all epochs, none of the shadows are fully stationary. This indicates that the atmosphere of the inner disk is a dynamical environment in which the gas and the micron-sized dust grains (which are dynamically coupled) are subjected to processes that change their vertical distribution on a fast timescale.

Variability of the shadows might be caused by orbital and/or vertical motion of dust enhancements in the inner disk. The vertical response of the disk occurs approximately on the Keplerian timescale meaning that a local perturbation of the disk will settle to an equilibrium state within approximately one orbit. However, the orbital timescale provides only a lower limit on the response timescale as it depends on the heating and cooling timescales of the gas. The $H$-band flux is emitted from a characteristic radius of 0.11\,au (see Section~\ref{sec:fit}) which corresponds to a Keplerian timescale of 10\,days, similar to the finest temporal sampling of the SPHERE imagery (8\,days). However, the shadows are presumably cast further outward as inferred from the absence of significant light travel time effects (see Section~\ref{sec:brightness_variations}). The temporal sampling of the SPHERE observations is too sparse to determine the timescale by which the shadow features appear and disappear, therefore, disentangling variability due to orbital motion and the (dis)appearance of shadows is not possible.

The $JHK$ fluxes in Figure~\ref{fig:photometry} show variations up to 10\% on a timescale of days to weeks \citep[see also][]{grady2009,sitko2012} while the $g'r'i'z'$ fluxes varied only by 1--2\%, indicating that mainly the thermal emission from the inner disk is affecting the near-infrared variability. For reference, the $J$-band flux of HD~135344B consists of 69\% stellar radiation and 31\% inner disk emission (see Section~\ref{sec:radiative_transfer}). The fast variability timescale of both the shadows and the near-infrared photometry may point toward a common origin in the inner disk.

Small variations of the visible photometry might be related to episodic accretion events, photospheric or chromospheric activity, stellar pulsations, minor attenuation variations by an (optically thin) dust envelope, or increased scattering from the inner disk. The scattered light flux from the inner disk is in the optical $\sim$1\% of the total flux so structural changes in the inner disk may cause a change both in thermal emission and scattered light. In that case, the photometric variations in the visible should be correlated with variations in the near-infrared which seems the case in the first half of the REM photometry but not the second half.

\subsubsection{Inner disk processes affecting the dust dynamics}\label{sec:processes}

Several processes may have an effect on the dynamics and distribution of the gas and dust in the inner disk, possibly causing shadow variations on the outer disk. For example, hydrodynamical fluctuations, such as turbulent eddies and filaments, may produce short-lived obscuration events \citep{dullemond2003,flock2017}. Or, catastrophic collisions between planetesimals, possibly stirred-up by a planet \citep{kenyon2004}, will locally enhance the dust density, although a gaseous environment will affect the dust dynamics differently than in a debris disk. Dedicated simulations are required to determine if such processes could produce disk perturbations that are localized and strong enough to explain the narrow shadows. Precession of a disk warp, for example driven by a companion \citep{lai2014}, will result in a variable location of the casted shadow \citep{debes2017}. The broad shadow in N-NW direction of HD~135344B appears approximately stationary so a fast precession can be excluded if the shadow is cast by a minor disk warp.

Variations in the inner disk might also be related to star-disk interactions. HD~135344B is an F4V-type star with a weak magnetic field \citep[$\langle B_{\rm z} \rangle = 32 \pm 15$\,G;][]{hubrig2009} so a magnetic coupling to and warping of the inner disk seems unlikely. Indeed, with the absence of a breaking mechanism, the star has been able to spin-up to a near break-up rotational velocity \citep{muller2011}. The fast rotation might drive a viscous decretion disk, similar to classical Be stars \citep{rivinius2013}, by which gas from the stellar atmosphere spreads outward, possibly creating disturbances in the inner disk by an interaction with the inward accretion flow. Also accretion may play a role in the distribution of material in the inner disk. \citet{fairlamb2015} measured a rate of $10^{-7.4}$\,$M_\odot$\,yr$^{-1}$ and \citet{sitko2012} determined a factor of two variation during the course of a few months. Therefore, accretion of gas and dust from the outer disk might be an irregular process, possibly mediated by one or multiple companions inside the large dust cavity, such that the inner disk gets asymmetrically replenished.

The near-infrared emission, hydrogen line fluxes, and the \ion{He}{1} line profile are all variable on various timescales, related to the processes occurring in the inner disk and near the stellar surface \citep{grady2009,sitko2012}. The variable P~Cygni profile of the \ion{He}{1} line is likely related to a wind whose orientation changes on timescales of a day or less \citep{sitko2012} which is launched in the star-disk interaction region \citep[e.g.,][]{edwards2009}. Micron-sized dust grains in the inner disk atmosphere could be entrained by a photoevaporative wind that is driven by the UV radiation of the star \citep{owen2011,hutchison2016} or dust could be uplifted by a centrifugally driven disk wind \citep{bans2012}. An extended low-density atmosphere could be supported by the magnetic field of the inner disk which may explain a large near-infrared excess and possible shadowing \citep{turner2014}. Alternatively, the central star could drive a wind from the circumplanetary disk of a planetary companion \citep{tambovtseva2006} or disk perturbations by a companion on an inclined orbit may also cause an asymmetric illumination of the disk \citep{demidova2013}. Three-dimensional radiation nonideal magnetohydrodynamical simulations show turbulent velocities in the inner disk up to 10\% of the sound speed and a nonaxisymmetric shadow on the outer disk cast by a dead zone-induced vortex \citep{flock2017}. An inner disk vortex will orbit the star with a Keplerian velocity so it is likely not responsible for the broad, quasi-stationary shadow.

\subsubsection{Face-on variant of the UX Orionis phenomenon?}\label{sec:uxor}

Although the origin of the shadows remains uncertain, we notice a resemblance between the photometric variations of UX~Orionis stars (UXORs) and the spatially resolved shadow variations on the disk surface around HD~135344B. UXORs are a subclass of Herbig Ae/Be stars which are characterized by sudden declines in brightness up to several magnitudes in the optical, associated with increased extinction and polarization, which suggests changes of the column density in the line of sight toward the star \citep{waters1998}. It has been proposed that such photometric variations could be caused by orbiting dust clouds when the disk is observed almost edge-on \citep{grinin1994} although alternative explanations involving disk winds \citep{grinin2003} and turbulent filaments exist \citep{dullemond2003}. Similar processes could be invoked to explain the variability of UXORs and HD~135344B, which may suggest that HD~135344B is a face-on variant of the UXOR phenomenon.

\subsection{Radiative transfer models of a shadowed outer disk}\label{sec:radiative_transfer}

\begin{figure}
\centering
\includegraphics[width=\linewidth]{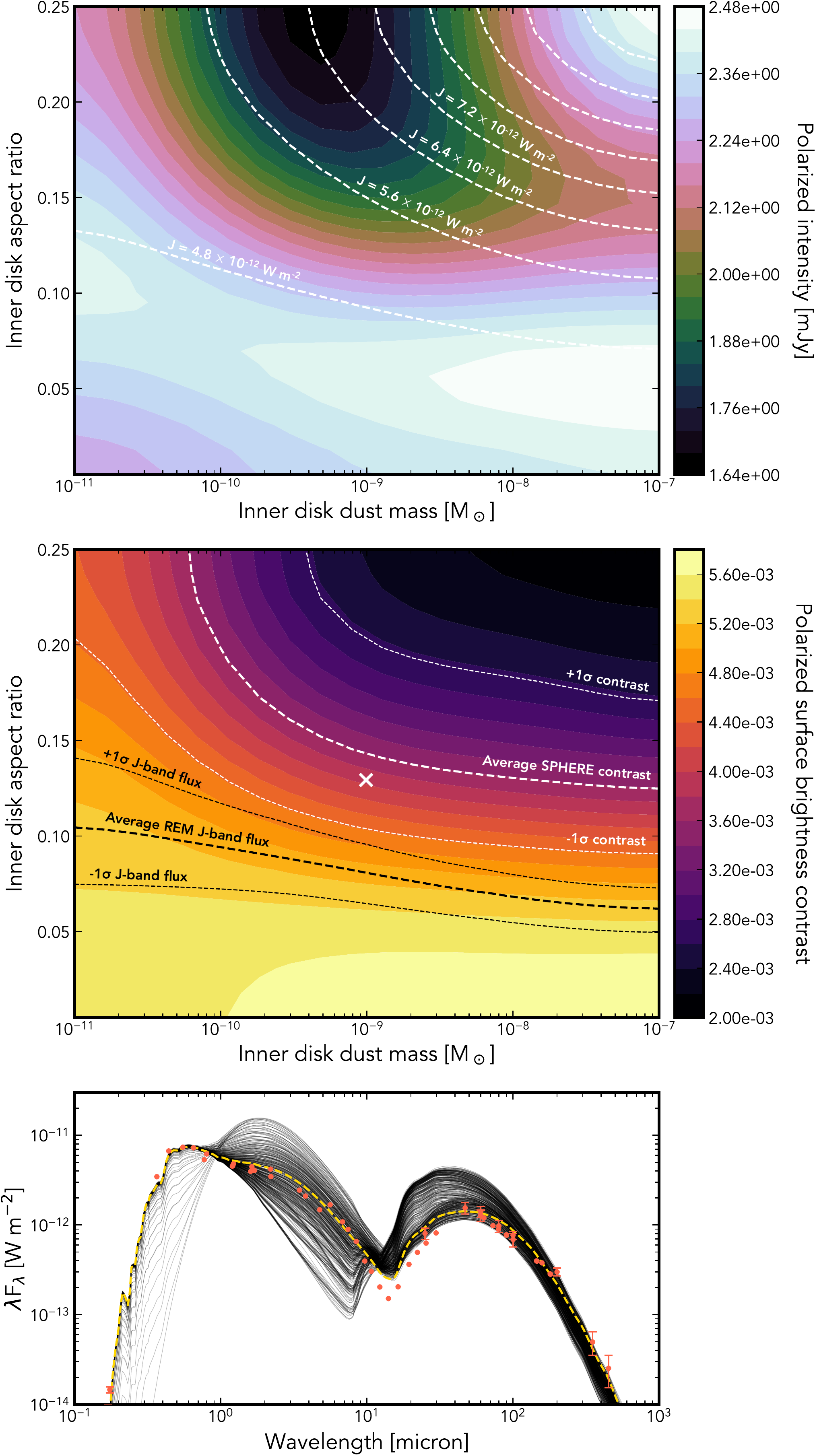}
\caption{\emph{Top}: radiative transfer simulations of the $Q_\phi$ intensity of the outer disk (colored map) and the $J$-band photometry (white contours). \emph{Center}: scattered light contrast (colored map) with the mean and 1$\sigma$ of the REM $J$-band flux (black dashed lines) and the contrast of the SPHERE imagery (white dashed lines). The white cross denotes to the DIANA model of HD~135344B \citep{woitke2016}. \emph{Bottom}: spectral energy distributions (SEDs) of all models (black solid lines), superimposed by SED of the DIANA model (yellow dashed line) and the photometry \citep[red points;][]{carmona2014}. Error bars of the photometry have been excluded when they are smaller than the symbol.\label{fig:mcmax}}
\end{figure}

Shadow variations on the outer disk depend on changes in the vertical distribution of dust in the inner disk. More specifically, the strength of the shadowing is set by the inner disk radius where the transition region from optically thin to thick occurs highest above the midplane. The height of this transition region is mainly determined by the pressure scale height and the surface density if the dust opacities are constant throughout the inner disk. For a flaring geometry, the thickest part will be close to the outer edge of the inner disk. Alternatively, a puffed-up inner rim may shadow the outer disk in case the exterior of the inner disk is fully shadowed by the rim \citep{dullemond2004,dong2015b}.

\subsubsection{Parametric model setup}\label{sec:model_setup}

To understand quantitatively the effect of the inner disk on the near-infrared excess and the scattered light contrast, without making an assumption about the origin of the shadows, we have constructed a grid of $20 \times 20$ radiative transfer models which are meant as a proof of concept rather than an accurate fit to the data. We adopted the DIANA model setup of HD~135344B which provides a multiwavelength fit to the SED and several other gas and dust observables \citep{woitke2016}. The inner disk is aligned with the outer disk and ranges from 0.16\,au to 0.21\,au (at 140\,pc), slightly beyond the characteristic radius inferred from the $H$-band visibilities (see Section~\ref{sec:fit}), with a surface density profile parameterized as $\Sigma \propto r^{-1.8}$ and a sharp inner rim. The pressure scale height profile is parameterized as $H\propto r^{-0.08}$ with the normalization provided by a reference aspect ratio, $H_0/r_0$, at $r_0=0.2$\,au. The negative flaring index implies a decrease of the scale height with increasing radius which is expected due to direct heating of the inner rim by the star. The radial extent of the inner disk is small so the flaring index has only a minor impact on the disk geometry. The grid covers values of the inner disk dust mass in the range of $10^{-11}$--$10^{-7}$\,$M_\odot$ and the inner disk aspect ratio, $H_0/r_0$, in the range of 0.005--0.25. The flaring index of the outer disk is 1.14 with a reference aspect ratio of 0.16 at 50\,au. The radiative transfer and image raytracing were done with \texttt{MCMax3D}, a Monte Carlo continuum radiative transfer code \citep{min2009}.

\subsubsection{Polarized scattered light versus thermal emission}\label{sec:scattered_thermal}

For each model, we computed the disk-integrated $Q_\phi$ intensity of the outer disk and the total $J$-band flux which are displayed in the top panel of Figure~\ref{fig:mcmax}. Several effects of the inner disk mass and the aspect ratio on the polarized intensity are evident in the colored map. Increasing the aspect ratio from very small values up to $\sim$0.05 results in a larger fraction of the stellar light being reprocessed by the inner disk and reemitted in the $J$ band, thereby increasing the scattered light flux from the outer disk. Similarly, the increase of the polarized intensity with increasing dust mass is also the result of a larger fraction of thermal radiation from the inner disk scattering from the outer disk. In this regime of the aspect ratio, the opening angle of the inner disk is too small to shadow the outer disk.

Shadowing of the outer disk by the inner disk starts to have an effect for $H_0/r_0\gtrsim0.05$ such that the opening angle of the inner disk atmosphere is comparable to or larger than the opening angle of the scattering surface of the outer disk. The polarized intensity decreases with increasing aspect ratio when the inner disk dust mass is smaller than $\sim$$10^{-9}$\,$M_\odot$ because the optical depth through the inner disk atmosphere toward the outer disk increases. Consequently, a larger fraction of the stellar light is attenuated by extinction in the inner disk. For inner disk masses larger than $\sim$$10^{-9}$\,$M_\odot$, the effect of enhanced reprocessing by the inner disk starts to dominate over the shadowing effect. More specifically, even though increasing the aspect ratio beyond 0.05 will increase the amount of shadowing of the outer disk, enhanced reprocessing by a larger amount of dust in the inner disk results in an increased irradiation of the outer disk in the $J$ band. Since the exponent is negative for both the surface density and the scale height profile, the inner disk is thickest at the location of the inner rim (i.e., the radius where the vertical $\tau=1$ surface is located highest above the midplane). As a result, reprocessed stellar light by the inner rim can directly irradiate the outer disk, in contrast to an inner rim which is partly shielded from the outer disk by a flaring scale height profile.

The $J$-band photometry of the radiative transfer models is superimposed on the polarized surface brightness in the top panel of Figure~\ref{fig:gallery}. The inner disk dust mass and aspect ratio have a similar effect on the $J$-band flux because both parameters affect the amount of reprocessing by the inner disk. While the stellar $J$-band flux remains constant, the thermal emission from the inner disk is positively correlated with both the inner disk dust mass and the reference aspect ratio. The SEDs of all models are shown in the bottom panel of Figure~\ref{fig:gallery}. As expected, the near- and mid-infrared excess varies greatly in the explored parameter space.

\subsubsection{Observed and simulated scattered light contrast}\label{sec:contrast}

An absolute flux calibration of the SPHERE data is not possible with high-precision so we use a relative brightness measurement to compare the results from the grid of models with the observations. The central panel of Figure~\ref{fig:mcmax} displays the polarized surface brightness contrast that has been computed with the same procedure described in Section~\ref{sec:brightness_variations}, here simply given by the ratio of the mean polarized intensity of the outer disk and the total $J$-band flux. The $J$-band flux and shadowing of the outer disk are interlinked in the lower part of the disk mass regime, therefore, the contrast distribution across the grid appears similar to the individual components in the top panel. In the regime of the large inner disk mass and aspect ratio, the contrast balances between the increasing scattered light flux and the increasing $J$-band photometry. The latter dominates over the former such that the contrast decreases toward the top right of the distribution, yet, the polarized intensity has a significant effect on the contrast. This means that using a relative flux measurement instead of an absolute one may lead to a degeneracy in the interpretation because a low contrast could correspond to either a high or low polarized intensity.

The azimuthally and temporally averaged scattered light contrast of the SPHERE observations, as well as the REM $J$-band photometry, are superimposed on the contrast from the radiative transfer models in the central panel of Figure~\ref{fig:mcmax}. We rejected the third quadrant because the brightness of the SW part of the disk is mainly determined by the enhanced midplane density which would otherwise bias the characteristic contrast variations by the shadowing. Significant changes in the total dust mass of the inner disk are not likely to occur as the local viscous timescale will be multiple orders of magnitude longer than the timescale probed with the SPHERE observations. For a fixed dust mass of $9.86 \times 10^{-10}$\,$M_\odot$ from the DIANA model, the observed scattered light contrast corresponds to aspect ratio values in the range of 0.11--0.22, likely beyond what is expected from a hydrostatically supported disk. The observed $J$-band fluxes on the other hand cover a lower aspect ratio regime (0.07--0.1). A less extended atmosphere will be required when a small misalignment is present between the inner and outer disk (see Section~\ref{sec:misalignment}) in which case the estimated value of $H_0/r_0$ will be smaller.

\subsubsection{Evidence for extended variations in the inner disk}\label{sec:extended_variations}

We infer from Figure~\ref{fig:mcmax} that extended variations in the inner disk atmosphere ($H/r \lesssim 0.2$) can explain the variations of the scattered light contrast and the related shadowing. However, the predicted $J$-band flux from an extended atmosphere is too large compared to the REM photometry. The discrepancy between the scattered light contrast and $J$-band flux possibly points to the uncertain origin of the near-infrared excess. We want to stress that the goal of the radiative transfer models is to provide a quantitative estimate of the required extent of the inner disk atmosphere without making an assumption about the origin of the near-infrared excess (e.g., super-refractory grains, disk wind, magnetically supported atmosphere). Nonetheless, alternative disk structures or dust properties could be considered to dissolve the discrepancy. For example, an inner disk consisting of an optically thick but geometrically thin component and an optically thin but geometrically extended component. The relative strength of the near-infrared excess and the outer disk shadowing is also sensitive to the abundances of amorphous silicates, amorphous carbon, and metallic iron. Amorphous laboratory silicates have a single scattering albedo close to unity \citep{dorschner1995} while the absorption cross section of carbon and iron grains is larger \citep{zubko1996}. Therefore, increasing the relative amount of silicate grains will lower the near-infrared excess while the extinction through the atmosphere may remain large enough to cast a shadow on the outer disk.

We also tested the dependence of the turbulence mixing strength, set by the dimensionless viscosity parameter $\alpha$, on the scattered light contrast and near-infrared excess. The parameter controls the dust settling in the radiative transfer models by assuming an equilibrium between upward turbulent mixing and downward gravitational settling \citep{woitke2016}. The turbulence/settling parameter mainly affects the brightness contrast when its value is $\lesssim$$10^{-4}$--$10^{-5}$ such that even the micron-sized grains slightly settle. The maximum $\alpha$ value affecting the brightness contrast increases with increasing aspect ratio. The dependence on the flaring index of the inner disk was also tested. As expected, for a given reference aspect ratio there appeared no significant dependence on the flaring index because of the narrow radial extent of the inner disk.

\section{Summary and conclusions}

We have presented a multi-epoch scattered light study of the protoplanetary disk surrounding HD~135344B, an isolated pre-main-sequence F4V-type star in the Scorpius \mbox{OB2-3} association. Polarimetric differential imaging observations with VLT/SPHERE in the $J$ band revealed, with a spatial resolution of $\sim$6.4\,au, azimuthal shadowing variations on the outer disk, related to the vertical dust distribution of dust in the inner disk. The imagery shows irregularly changing shadow patterns between all epochs although similarities have been identified. Shadows appear both as localized lanes and broader structures, typically colocated, which likely trace small scale perturbations and large scale dynamics, respectively. A broad, quasi-stationary shadow is present in N-NW direction of all scattered light images, in particular well visible at the cavity edge where the outer disk scattering surface is lowest. This might suggest that the inner disk is misaligned by several degrees, requiring additional optical depth enhancements through the inner disk atmosphere to shadow the disk further outward. However, fitting of a parametric brightness model to VLTI/PIONIER $H$-band visibilities provided a best-fit inclination and position angle which is, within the uncertainties, consistent with the outer disk. The photometry showed only minor variations (1--2\%) in the $g'r'i'z'$ bands while the $JHK$ fluxes varied up to 10\%, indicating significant changes in the amount of reprocessing of stellar light by the inner disk. Variability of the photometry and shadowing appear to be correlated, both are related to structural changes in the inner disk which we have quantified with a grid of radiative transfer models. The observed variations in scattered light contrast require extended variations in the inner disk atmosphere ($H/r \lesssim 0.2$), beyond what is inferred from the near-infrared excess alone, highlighting the uncertainty about the origin of the near-infrared excess. The variability of the shadows is likely related to the structure and dynamics of the inner disk, whose $H$-band emission originates from a characteristic radius of 0.11\,au, that is, inside the silicate sublimation zone. Asymmetric shadowing variations by the inner disk might be caused by mechanisms such as turbulent fluctuations, planetesimal collisions, a dusty disk wind, or asymmetric and episodic accretion. Simultaneous observations of scattered light, photometry, spectroscopy, and/or near-infrared interferometry will provide more stringent constraints on the driving processes in the inner disk for which an approximate daily sampling is required to trace the fast disk dynamics.

\acknowledgments

We would like to thank Doug Lin, Mario Flock, and the other attendees of the Ringberg workshop on `The Atmospheres of Disks and Planets' for insightful discussions. We also thank the referee for providing constructive comments and suggestions. This research has made use of the \texttt{OiDB}\footnote{Available at http://oidb.jmmc.fr} and \texttt{SearchCal}\footnote{Available at http://www.jmmc.fr/searchcal} services at the Jean-Marie Mariotti Center.

MB acknowledges funding from ANR of France under contract number ANR-16-CE31-0013 (Planet Forming Disks). SP acknowledges support from CONICYT-Gemini grant 32130007. SK acknowledges support from an STFC Rutherford fellowship (ST/J004030/1) and an ERC Starting Grant (Grant Agreement No. 639889).

\vspace{5mm}

\facilities{VLT:Melipal (SPHERE), VLTI (\mbox{PIONIER})}

\software{MCMax3D \citep{min2009}, SPHERE DRH \citep{pavlov2008}, SearchCal \citep{bonneau2006,bonneau2011}, Astropy \citep{astropy2013}}

\appendix

\section{Details on the visibility fitting}\label{sec:appendix}

The PIONIER $H$-band visibilities can be expressed as \citep{lazareff2017},
\begin{equation}
V(u,v,\lambda) = \frac{f_{\rm s}(\lambda_0/\lambda)^{k_{\rm s}}+V_{\rm c}(u,v)f_{\rm c}(\lambda_0/\lambda)^{k_{\rm c}}}{f_{\rm s}(\lambda_0/\lambda)^{k_{\rm s}}+f_{\rm c}(\lambda_0/\lambda)^{k_{\rm c}}},
\end{equation}
where $V_{\rm c}$ is the visibility of the circumstellar component, $k_{\rm s}$ and $k_{\rm c}$ are the spectral index of the stellar and circumstellar component, respectively, $f_{\rm s}$ and $f_{\rm c}$ are the fractional flux of the stellar and circumstellar component at a reference wavelength $\lambda_0$, and $\lambda$ is the wavelength of the spectral channel. The flux fractions of the star and circumstellar disk are normalized such that $f_{\rm s}+f_{\rm c}=1$.

As described in Section~\ref{sec:fit}, the visibilities are fitted with an ellipsoidal brightness distribution which is parameterized by a weighted combination of a Gaussian, $\mathcal{F}_{\rm G}(r)$, and a pseudo-Lorentzian, $\mathcal{F}_{\rm L}(r)$, radial distribution which allows for some freedom in the steepness of the asymptotic decay. The brightness distributions are derived from,
\begin{eqnarray}
\mathcal{F}_{\rm G}(r) & = \frac{\ln{2}}{\pi a^2}\exp{\left[ -\left(\frac{r}{a}\right)^2\ln{2} \right]}, \\
\mathcal{F}_{\rm L}(r) & = \frac{a}{2\pi\sqrt{3}} \left( \frac{a^2}{3} + r^2 \right)^{-3/2},
\end{eqnarray}
followed by an anamorphosis along the minor axis, and where $r$ is the emission radius and $a$ the semi-major axis of the half-light isophote. The Hankel transforms of the brightness distributions take a simple analytical form. The visibility of the circumstellar component is,
\begin{equation}
V_{\rm c} = (1 - f_{\rm L}) V_{\rm G} + f_{\rm L} V_{\rm L},
\end{equation}
where $f_{\rm L}$ is a weighting factor for the contribution of the pseudo-Lorentzian component, and $V_{\rm G}$ and $V_{\rm L}$ are the Hankel transforms of the Gaussian and pseudo-Lorentzian brightness distribution, respectively.

The best fit results in Section~\ref{sec:fit} were obtained by selecting the first quartile of the visibility data, on the basis of individual error estimates. Table~\ref{tab:interferometry} provides an overview of all the fitting results of the PIONIER visibilities to show the effect of the cutoff level on the best-fit values. The cutoff level in the cumulative distribution of the error estimates is stepwise loosened from 25\% up to 100\%, that is, the first quartile and all available data points, respectively. The error estimates for individual data points are derived by the PIONIER reduction pipeline from the internal dispersion on a short timescale. However, experience shows that other errors, not captured by the estimate, become increasingly prevalent as the observing conditions degrade. Our choice of the first quartile is a compromise between a severe selection that discards valid information and a loose selection that includes corrupted data, as is reflected in the increase of the reduced $\chi^2$ in Table~\ref{tab:interferometry}.

\begin{deluxetable*}{ccccccccccc}
\tablecaption{Visibility fitting results\label{tab:interferometry}}
\tabletypesize{\footnotesize}
\tablecolumns{10}
\tablewidth{0pt}
\tablehead{
\colhead{Cutoff} & \colhead{$n_{\rm V}$} & \colhead{$\chi^2_{\rm r}$} & \colhead{$i$} & \colhead{PA} & \colhead{$a$} & \colhead{$k_{\rm c}$} & \colhead{$f_{\rm c}$} & \colhead{$f_{\rm L}$} \\
\colhead{(\%)} & \colhead{} & \colhead{} & \colhead{(deg)} & \colhead{(deg)} & \colhead{(mas)} & \colhead{} & \colhead{} & \colhead{} & \colhead{} & \colhead{}
}
\startdata
 25 & 199 & 1.10 & $18.2^{+3.4}_{-4.1}$ & $57.3\pm5.7$  & $0.71\pm0.03$ & $-3.34\pm0.57$ & $0.61\pm0.04$ & $0.20\pm0.02$ \\
 35 & 278 & 1.44 & $19.9^{+3.1}_{-3.7}$ & $57.3\pm6.3$  & $0.76\pm0.03$ & $-3.04\pm0.57$ & $0.56\pm0.03$ & $0.17\pm0.03$ \\
 45 & 358 & 1.67 & $19.9^{+3.1}_{-3.7}$ & $56.1\pm5.7$  & $0.74\pm0.03$ & $-2.81\pm0.60$ & $0.57\pm0.04$ & $0.18\pm0.03$ \\
 55 & 437 & 1.85 & $21.6^{+2.9}_{-3.4}$ & $56.7\pm6.3$  & $0.74\pm0.05$ & $-2.79\pm0.68$ & $0.58\pm0.04$ & $0.18\pm0.03$ \\
 65 & 517 & 2.07 & $18.2^{+3.4}_{-4.1}$ & $56.7\pm7.4$  & $0.72\pm0.05$ & $-2.70\pm0.59$ & $0.58\pm0.04$ & $0.18\pm0.03$ \\
 75 & 597 & 2.28 & $19.9^{+3.1}_{-3.7}$ & $51.6\pm9.7$  & $0.78\pm0.05$ & $-2.64\pm0.62$ & $0.55\pm0.04$ & $0.15\pm0.03$ \\
 85 & 676 & 2.76 & $18.2^{+3.4}_{-4.1}$ & \phn$48.1\pm10.3$ & $0.74\pm0.05$ & $-2.39\pm0.70$ & $0.57\pm0.05$ & $0.18\pm0.03$ \\
 95 & 756 & 2.99 & $18.2^{+3.4}_{-4.1}$ & $45.3\pm9.1$  & $0.76\pm0.05$ & $-2.66\pm0.77$ & $0.57\pm0.06$ & $0.16\pm0.04$ \\
100 & 796 & 3.07 & $19.9^{+3.1}_{-3.7}$ & $49.8\pm9.7$  & $0.74\pm0.05$ & $-1.95\pm0.76$ & $0.58\pm0.05$ & $0.20\pm0.04$ \\
\enddata
\tablecomments{Table columns (from left to right): cutoff level in the cumulative distribution of the error estimates, number of $(u,v)$ points, reduced $\chi^2$, inclination, position angle of the major axis, half-flux semi-major axis, spectral index of the circumstellar component, fractional flux of the circumstellar component, and weighting factor of the pseudo-Lorentzian profile. The uncertainties are provided at a 1$\sigma$ level. Most parameter values appear stable over a broad range of quality cutoff levels.}
\end{deluxetable*}

\end{document}